\documentclass{bmvc2k}
\title{Prior-Guided Implicit Neural Representations for Single-Subject \\Diffusion MRI Super-Resolution}

\addauthor{Abdulkader Ghandoura$^{*}$}{aghandoura@bwh.harvard.edu}{1,2}
\addauthor{Marsil Zakour$^{\dagger}$}{marsil.zakour@tum.de}{2}
\addauthor{William Consagra$^{\ddagger}$}{consagra@mailbox.sc.edu}{1,3}
\addauthor{Yogesh Rathi$^{\ddagger}$}{yogesh@bwh.harvard.edu}{1}

\addinstitution{
 Psychiatry Neuroimaging Laboratory\\
 Brigham and Women's Hospital\\
 Harvard Medical School\\
 Boston, MA, USA
}
\addinstitution{
 School of Computation,\\
 Information and Technology\\
 Technical University of Munich\\
 Munich, Germany
}
\addinstitution{
 Department of Statistics\\
 University of South Carolina\\
 Columbia, SC, USA
}

\runninghead{Ghandoura \bmvaEtAl}{Prior-Guided INRs for dMRI Super-Resolution}

\usepackage{verbatim}
\usepackage{amssymb}
\usepackage{booktabs}
\usepackage{multirow}
\usepackage{makecell}

\makeatletter
\newcommand\blfootnote[1]{%
  \begingroup
  \renewcommand\thefootnote{}%
  \long\def\@makefntext##1{\noindent ##1}%
  \footnote{#1}%
  \addtocounter{footnote}{-1}%
  \endgroup
}
\makeatother
\begin{document}

\maketitle
\blfootnote{Accepted at the British Machine Vision Conference (BMVC) 2026.}
\blfootnote{$^{*}$Corresponding author. $^{\dagger}$M. Zakour is also with the Chair of Media Technology and Munich Institute of Robotics and Machine Intelligence, Technical University of Munich. $^{\ddagger}$These authors share senior authorship.}

\begin{abstract}

Resolving complex fiber geometries in brain white matter requires high-resolution diffusion MRI at the cost of long acquisition times. This leads many clinical protocols to opt for low-resolution scans, making downstream microstructure estimation and tractography challenging. Implicit neural representations (INRs) can model the diffusion signal continuously, enabling native single-subject super-resolution by querying the network at arbitrary spatial coordinates, yet existing methods often suffer from long training times and lack a mechanism to incorporate anatomical priors to regularize super-resolution by constraining the space of plausible reconstructions. To address these limitations, we propose a novel transfer-learning framework that pre-trains an INR on a high-resolution template and then adapts it to subject-specific scans via registration and fine-tuning. For \(4\times\) through-plane super-resolution from \(5\,\mathrm{mm}\) to \(1.25\,\mathrm{mm}\) on Human Connectome Project (HCP) data, our method reduces NRMSE by 36--49\% and increases FSIM by 24--43\% over a recent baseline with \(6\times\) faster training, outperforming competing INR-based methods across both image quality and domain-specific metrics. Code is available on the project page at \url{https://abdulkaderghandoura.github.io/research/msc-thesis/}.


\end{abstract}
\section{Introduction}
\label{sec:introduction}

Diffusion MRI (dMRI) is a non-invasive imaging technique that is used to probe the structure of brain tissue in vivo. The diffusion signal is related to the direction and magnitude of local water molecule diffusion, enabling the in vivo characterization of complex microstructural features and the tracing of large-scale white matter fiber pathways via tractography. Due to these capabilities, there is significant interest in using dMRI in the study of neurodegenerative diseases, neuropsychiatric disorders, and cognitive impairment~\cite{muller2024toward,jo2025white,zhao2025application}.
\par 
High-spatial-resolution diffusion MRI requires long acquisition times to achieve reasonable signal-to-noise ratios (SNR), pushing many clinical protocols toward thick-slice acquisitions to maintain SNR within feasible scanning durations, at the expense of through-plane spatial resolution. This loss in resolution causes standard reconstruction techniques to fail in recovering the fine-grained anatomical details needed for downstream analysis, such as tractography or microstructure estimation.
\par 
To address the limitations of microstructure inference and tractography in clinical acquisitions, we propose a super-resolution framework that reconstructs high-spatial-resolution dMRI signals from low-resolution clinical-grade scans. Building on recent advances in implicit neural representations (INRs) for dMRI \cite{consagra2024neural,dwedari2024estimating,hendriks2025implicit}, we parameterize the diffusion signal with a resolution-agnostic, orientation-consistent neural representation. The network is defined continuously over the image domain, naturally supporting super-resolution at arbitrary scales. Since super-resolution from very thick slices is severely ill-posed due to irreversible loss of detail, our proposed framework leverages transferable anatomical priors from a high-resolution template via registration-guided fine-tuning to estimate subject-specific INRs. This has the added benefit of amortizing computational cost through template pre-training, enabling fast super-resolution. Our main contributions are as follows.
\begin{itemize}
    \item A novel template-based transfer-learning framework that leverages anatomical priors via registration-guided fine-tuning for dMRI super-resolution.
    \item A demonstration of \(6\times\) faster training and superior perceptual quality compared to recent INR approaches on \(4\times\) through-plane super-resolution.
\end{itemize}

\section{Related Work}
\label{sec:related_work}

\subsection{Implicit Neural Representations}
Implicit neural representations (INRs), or neural fields, are neural network-based parameterizations of continuous functions that map spatial or spatiotemporal coordinates to corresponding field values, such as distance, color, or density. While INRs have been widely used in computer vision for 3D shape representation~\cite{park2019deepsdf}, novel view synthesis~\cite{mildenhall2021nerf}, and image super-resolution~\cite{chen2021learning}, their adoption in dMRI is relatively limited; \cite{consagra2024neural} employs a global SIREN-based~\cite{sitzmann2020implicit} INR to model single-shell dMRI and proposes an approach for predictive uncertainty quantification; \cite{hendriks2025implicit} integrates multi-shell constrained spherical deconvolution into an INR framework to continuously predict fiber orientation distribution functions (fODFs); \cite{dwedari2024estimating} employs multi-resolution hash encoding~\cite{muller2022instant}, a hierarchy of discrete grids with a compact MLP head, demonstrating significantly faster training compared to global parameterizations, but at the cost of gradient updates primarily affecting grid entries at supervised positions. All current INR-based methods train from scratch per subject and lack a mechanism for incorporating anatomical prior information.

\subsection{Diffusion MRI Super Resolution}
\label{ssec:dmri_sr}
Diffusion MRI data is collected as a 4D array $n_x \times n_y \times n_z \times M$, where $n_x,n_y,n_z$ are the number of grid points in the $X-Y-Z$ spatial dimensions, and $M$ is the number of diffusion-weighted volumes, each corresponding to a b-vector in q-space $p \in S^2$, indicating the direction of the applied magnetic gradient, and diffusion weighting b-value $b \geq 0$ related to the amplitude, duration, and separation of the applied diffusion-encoding gradient pulses. Therefore, dMRI super-resolution targets spatial or angular resolution, or both. 
\par
Most existing work focuses either on spatial super-resolution of parameter maps estimated from the raw dMRI signal, such as the diffusion tensor \cite{alexander2017image,tanno2017bayesian}, or on angular super-resolution of the raw dMRI signal \cite{elaldi2021equivariant,lyon2023spatio}. For downstream analyses, such as tractography, arbitrary-scale spatial super-resolution on the dMRI signal is desirable, since it enables streamline propagation at any chosen resolution. 
\par
Because INRs represent images as continuous functions of space, they natively enable “zero-shot” super-resolution via evaluation at arbitrary spatial coordinates, making them a natural choice for this task. INR-based approaches developed specifically for spatial dMRI super-resolution have so far been limited to supervised approaches \cite{wu2024csr,spears2024learning}, in which the INR is trained as a decoder alongside a traditional deep encoder. However, this reliance on supervision can be problematic in clinical settings, where pathology may induce a substantial distribution shift relative to the training data.

\section{Methodology}
\label{sec:methodology}
\begin{figure}[tbp]
  \centering
  \includegraphics[width=\textwidth]{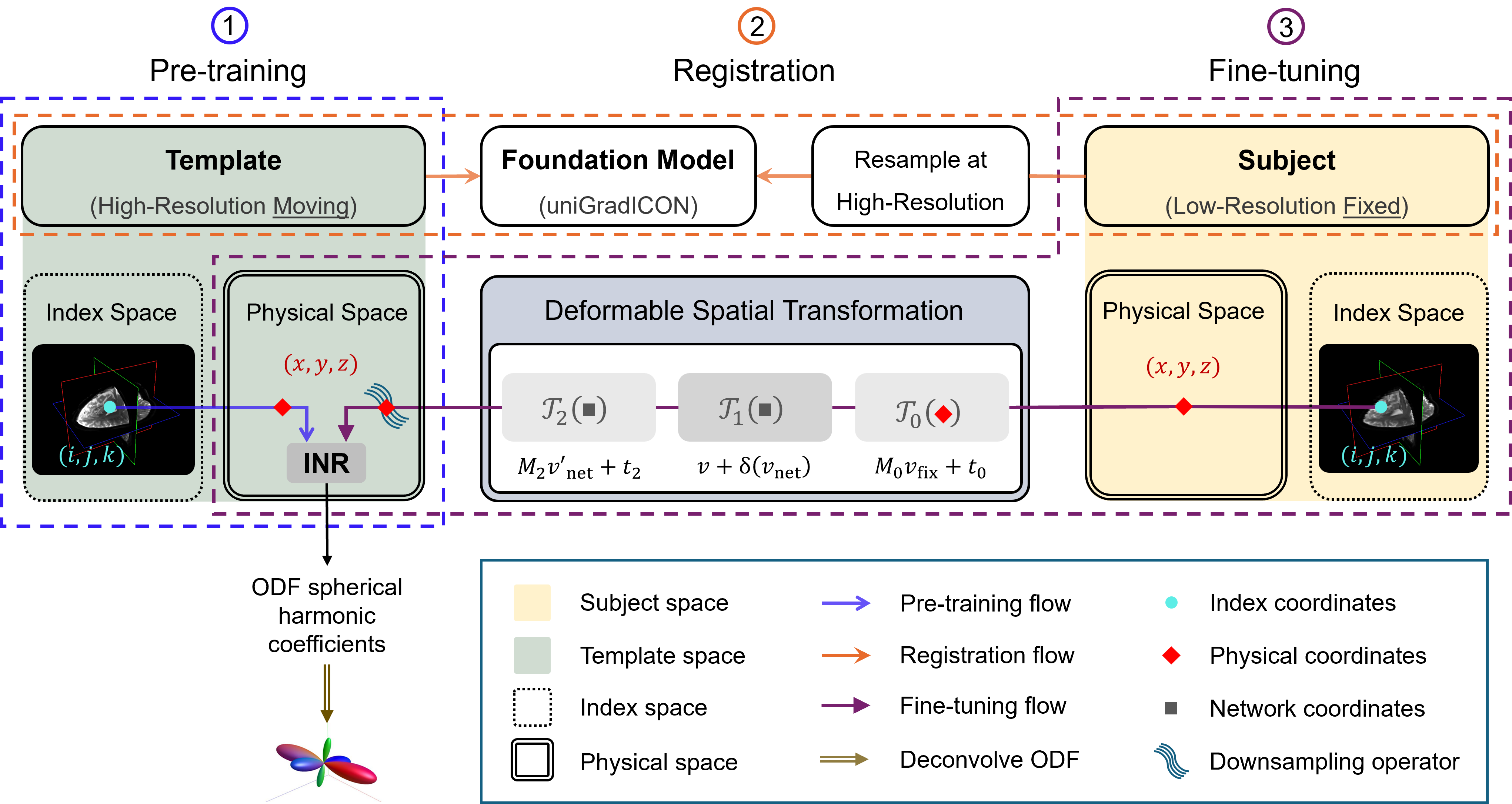} 
  \caption{Three-stage framework: (1) Pre-training an INR on a high-resolution dMRI template. (2) Subject-template registration via a composite transformation $\mathcal{T}$. (3) Fine-tuning to subject anatomy using low-resolution data.}
  \label{fig:pipeline}
\end{figure}

Figure~\ref{fig:pipeline} illustrates our prior-guided INR-based approach to single-subject dMRI super-resolution. In Stage 1 (Section~\ref{ssec:stage_1}), an INR is fit once to a high-resolution dMRI scan from a healthy subject, which will then serve as an anatomical template space prior for super-resolution on low-resolution subjects. In Stage 2 (Section~\ref{ssec:stage_2}), a non-rigid registration is performed to map spatial coordinates from the subject space to corresponding locations in the high-resolution prior template space. The subject-specific super-resolution is performed in Stage 3 (Section~\ref{ssec:stage_3}), where the pre-trained INR is injected into the subject-specific anatomy as a high-resolution prior and then fine-tuned to the new low-resolution data.

\subsection{High Resolution Template Space INR Prior}\label{ssec:stage_1}

Spherical harmonics provide a complete orthonormal basis for square-integrable functions on $S^2$. The complex spherical harmonics are defined as
\begin{equation}
Y_l^m(\theta, \varphi) = \sqrt{\frac{(2l + 1)(l - m)!}{4\pi(l + m)!}} P_l^m(\cos\theta) \, e^{im\varphi}
\end{equation}
for all non-negative integer orders \(l \geq 0\) and integer phase factors \(m = -l, \ldots, l\), where \(P_l^m\) are the associated Legendre polynomials. Due to the antipodal symmetry of diffusion, where water molecules diffuse equally in opposite directions, the ODF satisfies \(g_v(-p) = g_v(p)\) for all directions \(p \in S^2\). This symmetry property ensures that only even-order harmonics contribute to the ODF representation. 

The real-symmetric spherical harmonics form a set of basis functions \(\{\phi_k\}_{k=1}^K\) for antipodally symmetric functions, constructed from the complex harmonics as
\begin{equation}
\phi_k(p) = 
\begin{cases}
\sqrt{2} \, \mathrm{Re}(Y_l^m(p)) & \text{if } m < 0 \\
Y_l^0(p) & \text{if } m = 0 \\
\sqrt{2} \, \mathrm{Im}(Y_l^m(p)) & \text{if } m > 0
\end{cases}
\end{equation}
for even orders \(l = 0, 2, 4, \ldots, L\) and phase factors \(m = -l, \ldots, 0, \ldots, l\), where the index \(k\) is given by \(k = (l^2+l+2)/2 + m\). A key property is that these basis functions are non-zero eigenfunctions of the Funk-Radon transform, which relates the ODF to the diffusion signal, with eigenvalue \(2\pi P_l(0)\), where \(P_l\) is the ordinary Legendre polynomial of order \(l\).

We use the NODF model proposed in \cite{consagra2024neural} to parameterize the diffusion MRI signal, which is modeled as a continuous function of spatial location \(v \in \mathbb{R}^3\) and q-space orientation \(p \in S^2\) using the finite series expansion
\begin{equation}\label{eqn:dMRI_INR}
g(v, p) = \sum_{k=1}^{K} c_{\boldsymbol{\theta},k}(v) \phi_k(p),
\end{equation}
where \(\{\phi_k\}_{k=1}^K\) are the first $K$ real-symmetric spherical harmonic basis functions~\cite{descoteaux2007regularized}, and $\boldsymbol{c}_{\boldsymbol{\theta}}(v) = (c_{\boldsymbol{\theta},1}(v), \ldots, c_{\boldsymbol{\theta},K}(v))^\top$ is the corresponding spherical harmonic spatial coefficient field, parameterized by a SIREN-based INR \cite{sitzmann2020implicit}. 
\par 
Let \(\boldsymbol{\Phi} \in \mathbb{R}^{M \times K}\) be the real-symmetric spherical harmonic basis evaluation matrix at the \(M\) gradient directions with \(\Phi_{mk} = \phi_k(p_m)\), mapping the $K$-dimensional SH coefficient vector to the $M$-dimensional predicted diffusion signal along each gradient direction. We model the observed diffusion signal \(\mathbf{y}_i \in \mathbb{R}^M\) at voxel $i$ with center spatial coordinate \(v_i \in \mathbb{R}^3\) using a Gaussian measurement noise model with \(\boldsymbol{\epsilon}_i \sim \mathcal{N}(0, \sigma_e^2\boldsymbol{I}_{M})\) as
\begin{equation}
\mathbf{y}_i = \boldsymbol{\Phi} \boldsymbol{c}_{\boldsymbol{\theta}}(v_i) + \boldsymbol{\epsilon}_i.
\end{equation}
\par 
Training minimizes, over mini-batches of voxels of size~\(N\), a loss combining data fidelity (term 1) and an angular smoothness prior (term 2):
\begin{equation}\label{eqn:loss}
\widehat{\boldsymbol{\theta}}_{HR}
=\operatorname*{argmin}_{\boldsymbol{\theta}}\frac{1}{N} \sum_{i=1}^N \left( \|\mathbf{y}_i - \boldsymbol{\Phi} \boldsymbol{c}_{\boldsymbol{\theta}}(v_i)\|_2^2 + \lambda_c \,\boldsymbol{c}_{\boldsymbol{\theta}}(v_i)^\top \boldsymbol{R}_\gamma \boldsymbol{c}_{\boldsymbol{\theta}}(v_i) \right),
\end{equation}
where \(\boldsymbol{R}_\gamma \in \mathbb{R}^{K \times K}\) is the diagonal precision 
matrix encoding the angular prior (Section 3.1 of~\cite{consagra2024neural}), with 
\([\boldsymbol{R}_\gamma^{-1}]_{kk} = s_\gamma\bigl(\sqrt{l_k(l_k+1)}\bigr)\) for 
$s_\gamma$ the spectral density of the spherical Mat\'{e}rn prior with 
parameters~$\gamma$ and for $l_k$ the harmonic order of $\phi_k$, and \(\lambda_c\) 
controls the strength of the angular prior regularization.

\subsection{High-Resolution Template Prior Injection to Subject Space}\label{ssec:stage_2}
To inject the high-resolution INR-based prior into the low-resolution subject space for super-resolution, we first register the high-resolution template (moving) to the low-resolution subject (fixed) using registration foundation model \textit{uniGradICON}~\cite{tian2024unigradicon} with the local normalized cross-correlation (LNCC) similarity metric. This establishes a subject-to-template mapping, following the ITK conventions~\cite{avants2014insight}. Briefly, uniGradICON first resamples both volumes to $175^3$ voxels via trilinear interpolation, and then refines the alignment by minimizing $1 - \mathrm{LNCC}$. The resulting transformation maps fixed-space point $\tilde{v} \in \mathbb{R}^3$ to its moving-space correspondence $v$ through three stages. First, an affine $\mathcal{T}_0$ transforms to network space: $v_{\mathrm{net}} = \mathbf{M}_0 \tilde{v} + \mathbf{t}_0$. Second, the displacement vector field $\mathcal{T}_1$ is applied: $v_{\mathrm{net}}' = v_{\mathrm{net}} + \delta(v_{\mathrm{net}})$, where $\delta:\mathbb{R}^3 \rightarrow \mathbb{R}^3$ is interpolated in network space. Third, an affine $\mathcal{T}_2$ maps to moving space: $v = \mathbf{M}_2 v_{\mathrm{net}}' + \mathbf{t}_2$. The complete transform is $\mathcal{T} = \mathcal{T}_2 \circ \mathcal{T}_1 \circ \mathcal{T}_0$. 

\begin{figure}[!t]
  \centering
  \includegraphics[width=0.8\textwidth]{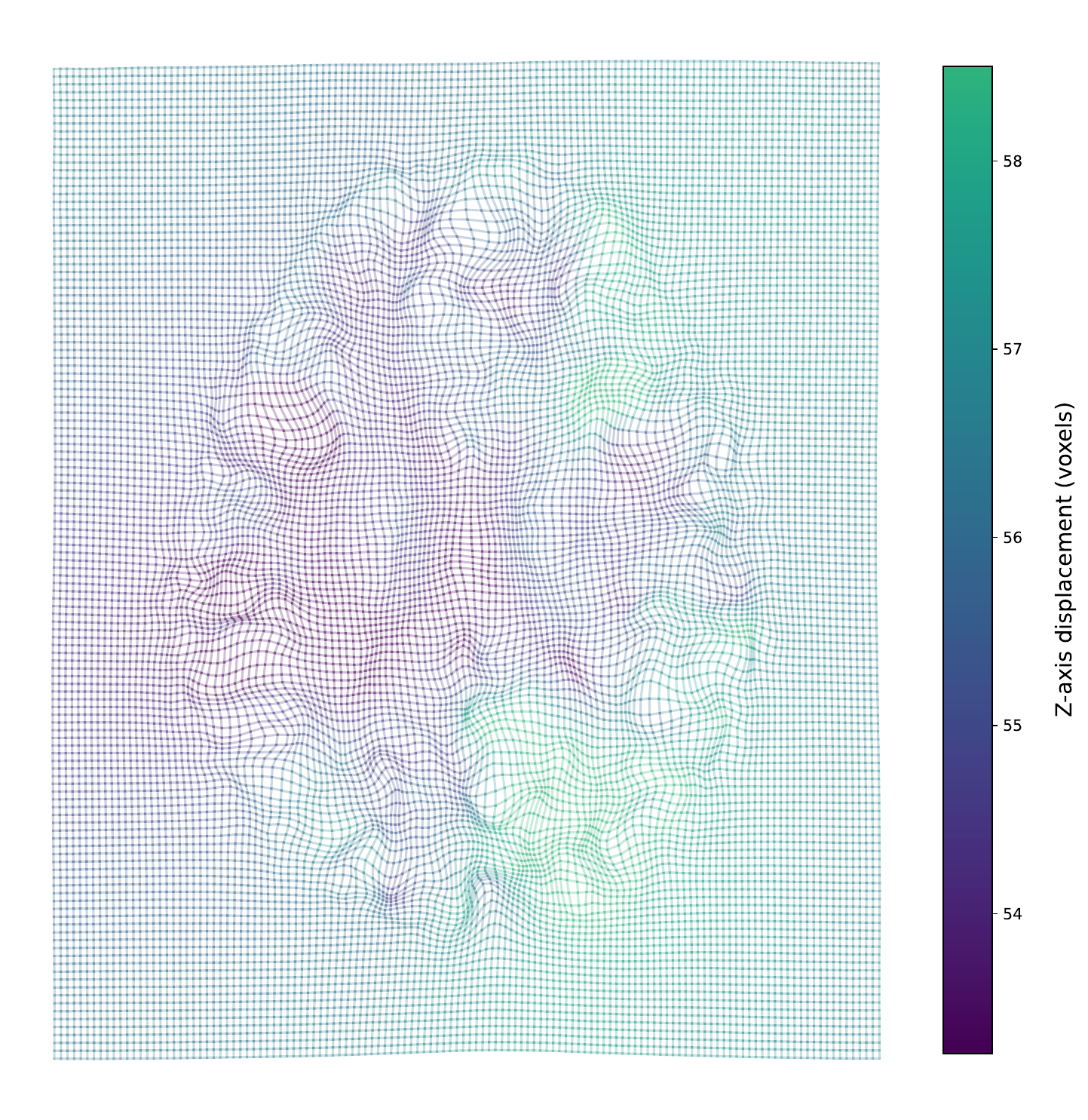} 
  \caption[Deformed grid showing through-plane displacement]{Deformed grid of an axial slice after applying the displacement field. A regular $128\times128$ grid created in the fixed image space was transformed through $\mathcal{T}$, resulting in curved grid lines that visualize the spatial deformation. Grid line colors encode the z-axis displacement of voxels, revealing a through-plane deformation pattern for a random subject.}
  \label{fig:axial_grid}
\end{figure}

Before registration, we used~\cite{consagra2024neural} with the hyperparameters proposed in Section~\ref{sec:experimental_setup} to resample the subject scan at high resolution. We then obtained the generalized fractional anisotropy (GFA) for registration to focus on white-matter structures, though alternatives, such as non-diffusion-weighted images~(\(b=0 \ s/mm^2\)) and fractional anisotropy (FA), are also suitable. Figure~\ref{fig:axial_grid} illustrates the resulting deformation for a randomly selected subject: $\mathcal{T}$ displaces points within and across the axial plane, reflecting the full three-dimensional anatomical realignment required between the subject and the template coordinate systems.

\subsection{Fine Tuning on Low-Resolution Subject}\label{ssec:stage_3}
We now introduce a transfer-learning formulation that uses the high-resolution template INR from Section~\ref{ssec:stage_1} together with the foundation model from Section~\ref{ssec:stage_2} to form a prior for single-subject super-resolution. The model is fine-tuned to enforce data consistency with the measured low-resolution signal, enabling single-subject super-resolution at arbitrary target voxel sizes. Specifically, for subject space voxel $i$, denote its spatial footprint \(\tilde{\Omega}_i\). Mapping points to template space via the deformable transform, we approximate the downsampling operator by Monte-Carlo integration 
\begin{equation}\label{eqn:ds_operator}
\tilde{\boldsymbol{c}}_{i} = \frac{1}{|\tilde{\Omega}_i|}\sum_{\tilde{v}^{(i)}\in \tilde{\Omega}_i} \boldsymbol{c}_{\boldsymbol{\theta}}\!\left(\mathcal{T}(\tilde{v}^{(i)})\right)\in \mathbb{R}^K,
\end{equation}
where $\tilde{v}^{(i)}$ are sampled uniformly over $\tilde{\Omega}_i$. Although our experiments use uniform slice averaging, the operator extends to any slice-excitation profile as a non-uniform weighting over the voxel footprint, with uniform averaging as the special case. For a mini-batch of $N$ voxels with $S$ samples per voxel, evaluating \eqref{eqn:ds_operator} requires $NS$ forward passes through $\boldsymbol{c}_{\boldsymbol{\theta}}$, i.e., $S{\times}$ the per-iteration cost of single-point evaluation at transformed voxel centers. To bound this overhead, we precompute a dense set of deformed footprint points $\mathcal{T}(\tilde{v}^{(i)})$ once at initialization and randomly draw $S$ of these points per voxel at each iteration, bounding GPU memory while preserving accuracy (Section~\ref{sec:ablation}). Let $\tilde{\mathbf{y}}_i \in \mathbb{R}^M$ denote the observed low-resolution diffusion signal at voxel $i$. We fine-tune the network parameters $\boldsymbol{\theta}$ by gradient updates according to the loss over mini-batches of $N$ voxels:
\begin{equation}\label{eqn:LR_training_objective}
    \mathcal{L}(\boldsymbol{\theta}) = \frac{1}{N}\sum_{i=1}^{N} \Big(
    \left\|\tilde{\mathbf{y}}_i - \boldsymbol{\Phi}\,\tilde{\boldsymbol{c}}_i\right\|_2^2
    + \lambda_c \,\tilde{\boldsymbol{c}}_i^\top \boldsymbol{R}_\gamma \tilde{\boldsymbol{c}}_i
    + \lambda_{\mathrm{tv}}\left\|\nabla_{\mathbb{R}^3}\boldsymbol{c}_{\boldsymbol{\theta}}\right\|_1
    \Big),
\end{equation}
where \(\lambda_{\mathrm{tv}}\) controls the strength of the spatial total-variation (TV) penalty, which is included here to regularize the spatial downsampling. Optimization of \eqref{eqn:LR_training_objective} is performed with Adam initialized at $\widehat{\boldsymbol{\theta}}_{HR}$, thereby injecting the high-resolution template space prior. Arbitrary-scale spatio-angular super-resolution at any subject location $\tilde{v}$ is then obtained via \eqref{eqn:dMRI_INR} using the fine-tuned network parameters.

\section{Experimental Setup}
\label{sec:experimental_setup}

\noindent{\textbf{Data}}: We used the publicly available HCP-YA dataset~\cite{van2013wu,sotiropoulos2013advances} with $1.25^3~\mathrm{mm}$ isotropic voxels acquired in a three-shell scheme with 90 directions per shell and 18 non-diffusion-weighted images. The dMRI data were processed with the HCP minimal preprocessing pipeline~\cite{glasser2013minimal,sotiropoulos2013advances}. For all experiments, we used only the $b = 1{,}000 \ \mathrm{s/mm^2}$ shell for high SNR. We randomly selected one subject as the high-resolution scan for training the template-space INR as outlined in Section~\ref{ssec:stage_1}. To mimic common clinical protocols with low out-of-plane resolution, we randomly selected 10 test HCP subjects and generated anisotropic low-resolution by averaging every 4 consecutive axial slices, yielding $1.25 \times 1.25 \times 5 \ \mathrm{mm}$ input voxels ($4\times$ through-plane reduction). The original high-resolution isotropic test volumes served as ground truth (GT) for evaluation.
\par 
\noindent{\textbf{Competing Methods}}:
To evaluate single-subject dMRI spatial super-resolution, we compare our method with two recent representative INR-based baselines: NODF \cite{consagra2024neural} and NODF-HashEnc \cite{dwedari2024estimating}. Both are trained using the default configuration reported in the latter. Because these models represent the signal as a continuous function of space, they support arbitrary-scale spatial super-resolution by querying the trained network at any spatial coordinate. Supervised CNN-based methods are excluded as spatial super-resolution methods for the dMRI signal remain scarce~\cite{alexander2017image,tanno2017bayesian}, and none support arbitrary-scale upsampling, which is essential for tractography at any chosen resolution. Moreover, such multi-subject supervision is problematic under the clinical distribution shift discussed in Section~\ref{ssec:dmri_sr}. In contrast, our template is a single anatomical prior, injected by registration and fully adapted to the subject's own data, involving no cross-subject supervision.
\par 
\noindent{\textbf{Evaluation Metrics}}: 
We use the ground-truth high-resolution data to compare the methods' recovery of microstructural features and orientation content from the low-resolution test images. For microstructure, we compare spatial feature maps of GFA and fractional anisotropy (FA), using standard image quality metrics: NRMSE, FSIM, PSNR, and SSIM. For orientational content, we compare the ODF $L_2$ error and the angular error of the principal eigenvector of the diffusion tensor. We also perform qualitative comparisons of tractography from the super-resolved estimates vs. tractography on the ground truth data.  
\par 
\noindent{\textbf{Implementation Details}}: Our template space INR uses a SIREN architecture with the same configuration as outlined above for the NODF competitor. We set $10$ fully-connected layers, uniform width $r=1{,}024$, mini-batch $N=128$, learning rate $10^{-6}$, $\lambda_c = 3.76 \times 10^{-7}$, and $\omega_0 = 60$ for both template training \eqref{eqn:loss} and subject-specific fine-tuning \eqref{eqn:LR_training_objective}. For the latter, we apply total variation regularization $\lambda_{tv}=2 \times 10^{-3}$ and use $S=6$ Monte Carlo samples per voxel in \eqref{eqn:ds_operator}. Training for all models was performed on an NVIDIA GeForce GTX~1080~Ti GPU with 11 GB VRAM. Image registration was carried out on a separate NVIDIA RTX A6000 GPU with 48 GB VRAM. The network has 10.5 million trainable parameters, yielding 42\, MB of model weights. As each method requires independent optimization for each test subject, the training time directly reflects the per-subject deployment cost. The template-space INR converges to a faithful reconstruction of the high-resolution input signal. At completion of training, the FA and GFA maps derived from the fitted INR are visually indistinguishable from those computed directly on the raw HCP template using DIPY~\cite{garyfallidis2014dipy}, with an NRMSE of 0.0349 on FA and 0.0782 on GFA, confirming that Stage~1 (Section~\ref{ssec:stage_1}) provides a high-fidelity anatomical prior for subsequent registration and fine-tuning.

\section{Results}
\subsection{Quantitative and Qualitative Evaluation}
\label{ssec:quantitative_and_qualitative_evaluation}
\begin{table}[!t]
\caption{
Quantitative results across training time on 10 randomly selected HCP subjects. Our method achieves the lowest NRMSE and the highest FSIM with consistent improvements for PSNR and SSIM, particularly on FA maps. Mean~\(\pm\)~standard deviation.
}
\renewcommand{\arraystretch}{1.2}
\centering
\footnotesize
\begin{tabular}{c c c r r r}
\toprule
\textbf{Metric} & \textbf{Map} & \textbf{Method} & \textbf{5 minutes} & \textbf{10 minutes} & \textbf{30 minutes} \\
\midrule
\multirow{6}{*}{\makecell{\textbf{NRMSE} $\downarrow$}} 
& \multirow{3}{*}{FA} 
& Baseline & 0.283 {\textcolor{gray}{$\pm$0.0031}} & 0.279 {\textcolor{gray}{$\pm$0.0034}} & 0.216 {\textcolor{gray}{$\pm$0.0114}} \\
& & HashEnc & 0.283 {\textcolor{gray}{$\pm$0.0046}} & 0.228 {\textcolor{gray}{$\pm$0.0309}} & 0.182 {\textcolor{gray}{$\pm$0.0044}} \\
& & Proposed & \textbf{0.149} {\textcolor{gray}{$\pm$0.0028}} & \textbf{0.143} {\textcolor{gray}{$\pm$0.0031}} & \textbf{0.138} {\textcolor{gray}{$\pm$0.0029}} \\
\cmidrule(lr){2-6}
& \multirow{3}{*}{GFA} 
& Baseline & 0.193 {\textcolor{gray}{$\pm$0.0398}} & 0.194 {\textcolor{gray}{$\pm$0.0394}} & 0.150 {\textcolor{gray}{$\pm$0.0280}} \\
& & HashEnc & 0.206 {\textcolor{gray}{$\pm$0.0408}} & 0.161 {\textcolor{gray}{$\pm$0.0272}} & 0.131 {\textcolor{gray}{$\pm$0.0254}} \\
& & Proposed & \textbf{0.104} {\textcolor{gray}{$\pm$0.0187}} & \textbf{0.100} {\textcolor{gray}{$\pm$0.0180}} & \textbf{0.096} {\textcolor{gray}{$\pm$0.0168}} \\
\midrule
\multirow{6}{*}{\makecell{\textbf{FSIM} $\uparrow$}} 
& \multirow{3}{*}{FA} 
& Baseline & 0.430 {\textcolor{gray}{$\pm$0.0090}} & 0.432 {\textcolor{gray}{$\pm$0.0105}} & 0.508 {\textcolor{gray}{$\pm$0.0139}} \\
& & HashEnc & 0.473 {\textcolor{gray}{$\pm$0.0250}} & 0.551 {\textcolor{gray}{$\pm$0.0351}} & 0.612 {\textcolor{gray}{$\pm$0.0038}} \\
& & Proposed & \textbf{0.606} {\textcolor{gray}{$\pm$0.0046}} & \textbf{0.617} {\textcolor{gray}{$\pm$0.0053}} & \textbf{0.630} {\textcolor{gray}{$\pm$0.0045}} \\
\cmidrule(lr){2-6}
& \multirow{3}{*}{GFA} 
& Baseline & 0.442 {\textcolor{gray}{$\pm$0.0084}} & 0.437 {\textcolor{gray}{$\pm$0.0167}} & 0.468 {\textcolor{gray}{$\pm$0.0225}} \\
& & HashEnc & 0.436 {\textcolor{gray}{$\pm$0.0172}} & 0.517 {\textcolor{gray}{$\pm$0.0481}} & 0.581 {\textcolor{gray}{$\pm$0.0183}} \\
& & Proposed & \textbf{0.579} {\textcolor{gray}{$\pm$0.0149}} & \textbf{0.588} {\textcolor{gray}{$\pm$0.0116}} & \textbf{0.598} {\textcolor{gray}{$\pm$0.0130}} \\

\midrule
\multirow{6}{*}{\makecell{\textbf{PSNR (dB)} $\uparrow$}} 
& \multirow{3}{*}{FA} 
& Baseline & 20.453 {\textcolor{gray}{$\pm$0.8323}} & 19.939 {\textcolor{gray}{$\pm$1.2143}} & 21.267 {\textcolor{gray}{$\pm$0.5898}} \\
& & HashEnc & 19.502 {\textcolor{gray}{$\pm$1.8340}} & 21.159 {\textcolor{gray}{$\pm$1.0218}} & 22.338 {\textcolor{gray}{$\pm$0.5027}} \\
& & Proposed & \textbf{22.963} {\textcolor{gray}{$\pm$0.3626}} & \textbf{23.278} {\textcolor{gray}{$\pm$0.3682}} & \textbf{23.627} {\textcolor{gray}{$\pm$0.3835}} \\
\cmidrule(lr){2-6}
& \multirow{3}{*}{GFA} 
& Baseline & \textbf{23.793} {\textcolor{gray}{$\pm$1.5542}} & 22.454 {\textcolor{gray}{$\pm$2.2521}} & 22.255 {\textcolor{gray}{$\pm$2.0966}} \\
& & HashEnc & 22.364 {\textcolor{gray}{$\pm$1.9013}} & 22.779 {\textcolor{gray}{$\pm$1.6999}} & 23.208 {\textcolor{gray}{$\pm$1.8943}} \\
& & Proposed & 23.151 {\textcolor{gray}{$\pm$2.6809}} & \textbf{23.019} {\textcolor{gray}{$\pm$2.3827}} & \textbf{23.268} {\textcolor{gray}{$\pm$2.1518}} \\
\midrule
\multirow{6}{*}{\makecell{\textbf{SSIM} $\uparrow$}} 
& \multirow{3}{*}{FA} 
& Baseline & 0.737 {\textcolor{gray}{$\pm$0.0181}} & 0.734 {\textcolor{gray}{$\pm$0.0206}} & 0.751 {\textcolor{gray}{$\pm$0.0147}} \\
& & HashEnc & 0.745 {\textcolor{gray}{$\pm$0.0190}} & 0.769 {\textcolor{gray}{$\pm$0.0175}} & 0.803 {\textcolor{gray}{$\pm$0.0112}} \\
& & Proposed & \textbf{0.807} {\textcolor{gray}{$\pm$0.0116}} & \textbf{0.816} {\textcolor{gray}{$\pm$0.0105}} & \textbf{0.830} {\textcolor{gray}{$\pm$0.0103}} \\
\cmidrule(lr){2-6}
& \multirow{3}{*}{GFA} 
& Baseline & \textbf{0.773} {\textcolor{gray}{$\pm$0.0244}} & 0.747 {\textcolor{gray}{$\pm$0.0351}} & 0.739 {\textcolor{gray}{$\pm$0.0230}} \\
& & HashEnc & 0.740 {\textcolor{gray}{$\pm$0.0259}} & 0.748 {\textcolor{gray}{$\pm$0.0237}} & 0.763 {\textcolor{gray}{$\pm$0.0283}} \\
& & Proposed & 0.761 {\textcolor{gray}{$\pm$0.0399}} & \textbf{0.760} {\textcolor{gray}{$\pm$0.0321}} & \textbf{0.767} {\textcolor{gray}{$\pm$0.0302}} \\
\bottomrule
\end{tabular}%
\label{tbl:metrics}
\end{table}

\begin{figure}[htbp]
  \centering
  \includegraphics[width=\textwidth]{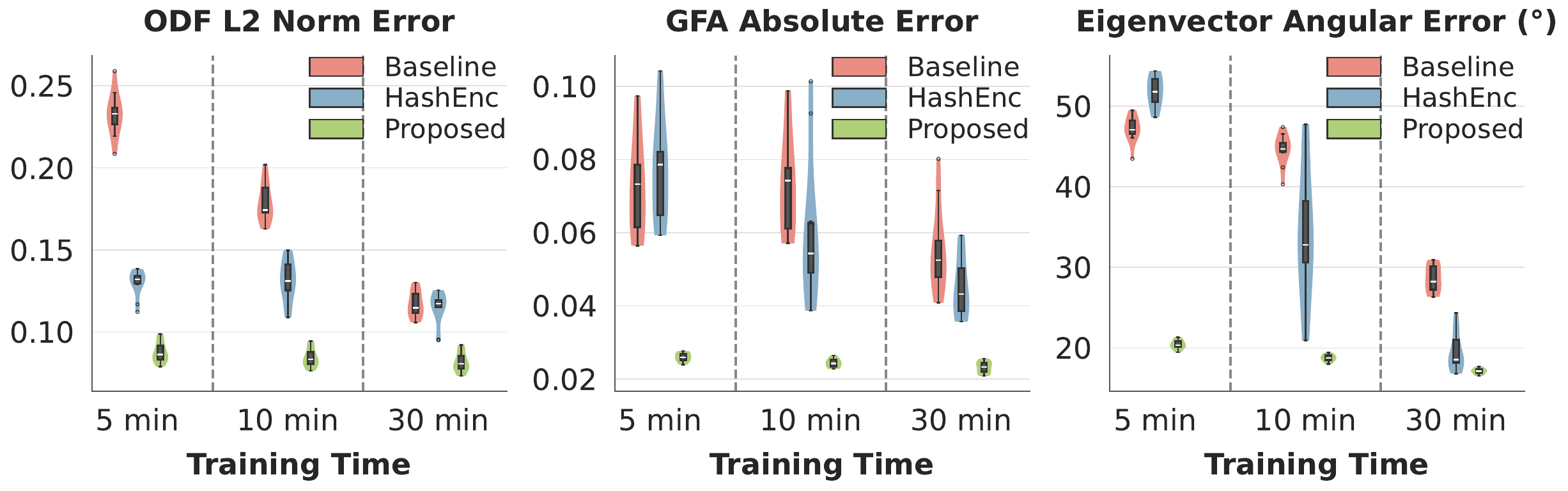} 
    \caption{Distribution of subject-wise median on 10 HCP subjects at 5, 10, and 30 minutes for ODF L2 norm error (Proposed 0.08--0.09 vs. Baseline (NODF) 0.11--0.23, HashEnc 0.12--0.13), GFA absolute error (0.023--0.026 vs. Baseline/HashEnc 0.04--0.08), and eigenvector angular error (17--20$^{\circ}$ vs. Baseline 28--47$^{\circ}$, HashEnc 19--52$^{\circ}$).}
  \label{fig:cv-median-metrics}
\end{figure}
\par 
Table~\ref{tbl:metrics} reports the results for the FA and GFA errors averaged over 10 test subjects for all methods as a function of training time, reflecting the practical per-subject deployment cost. Strikingly, our method achieves better performance in 5 minutes than the competing methods NODF and NODF-HashEnc even after 30 minutes of training, reducing NRMSE by 36--49\% and increasing FSIM by 24--43\% compared to NODF, while improving PSNR by 11--17\% and SSIM by 9--11\% for FA, with modest GFA gains at longer training budgets (see~\cite{breger2025study,mason2019comparison} for well-documented PSNR and SSIM limitations in medical imaging). We attribute this to the template-space prior introduced by our transfer-learning mechanism. Figure~\ref{fig:cv-median-metrics} summarizes the distributions of subject-wise median errors across three diffusion metrics on the test subjects using violin plots. Our method achieves lower ODF reconstruction error and smaller leading-diffusion-eigenvector angular error than both NODF and NODF-HashEnc, while converging substantially faster. The low inter-subject variability observed across all metrics (Table~\ref{tbl:metrics} and Figure~\ref{fig:cv-median-metrics}) indicates that these gains are consistent across subjects. To formally assess significance, we ran a two-sided paired Wilcoxon signed-rank test across the 10 subjects, Holm-corrected within each training-time budget. Our method significantly outperforms both baselines on all seven primary metrics (FA/GFA NRMSE, FA/GFA FSIM, ODF-L2, GFA absolute error, and angular error) at every budget ($\alpha=0.05$), and FA PSNR and FA SSIM are likewise significant.
\par
\begin{figure*}[t]
    \centering
    
    \hspace*{\dimexpr0.05\textwidth+1mm\relax}%
    \begin{minipage}[c]{0.188\textwidth}
        \centering\small Input\\(low-res)
    \end{minipage}
    \begin{minipage}[c]{0.188\textwidth}
        \centering\small Baseline\\($1\,\mathrm{h}$)
    \end{minipage}
    \begin{minipage}[c]{0.188\textwidth}
        \centering\small HashEnc\\($1\,\mathrm{h}$)
    \end{minipage}
    \begin{minipage}[c]{0.188\textwidth}
        \centering\small Proposed\\($5\,\mathrm{min}$ + init.)
    \end{minipage}
    \begin{minipage}[c]{0.188\textwidth}
        \centering\small Ground Truth\\(isotropic)
    \end{minipage}
    
    \begin{minipage}[c]{0.05\textwidth}
        \centering
        \rotatebox[origin=c]{90}{\small DTI}
    \end{minipage}
    \begin{minipage}[c]{0.188\textwidth}
        \centering
        \includegraphics[width=\textwidth]{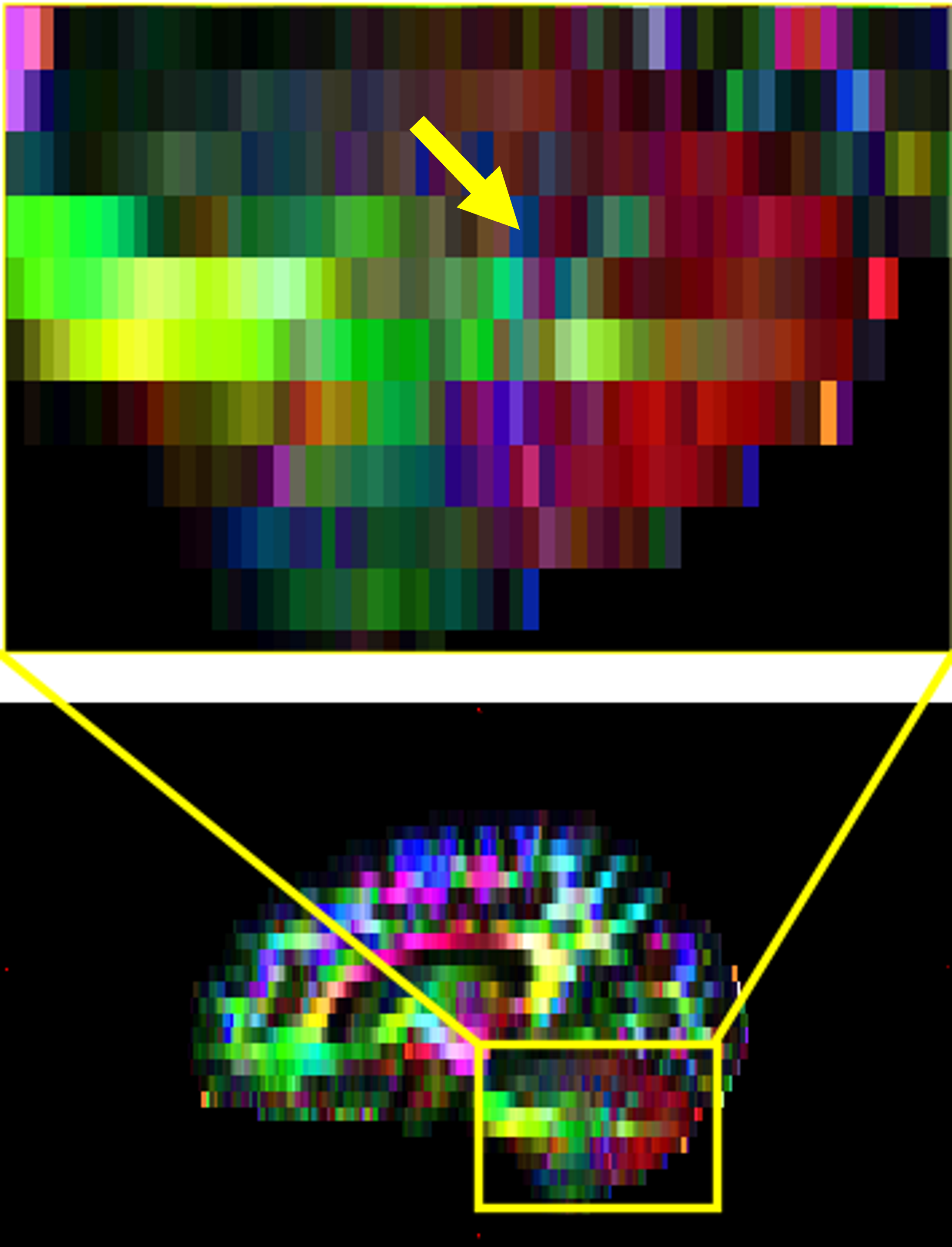}
    \end{minipage}
    \begin{minipage}[c]{0.188\textwidth}
        \centering
        \includegraphics[width=\textwidth]{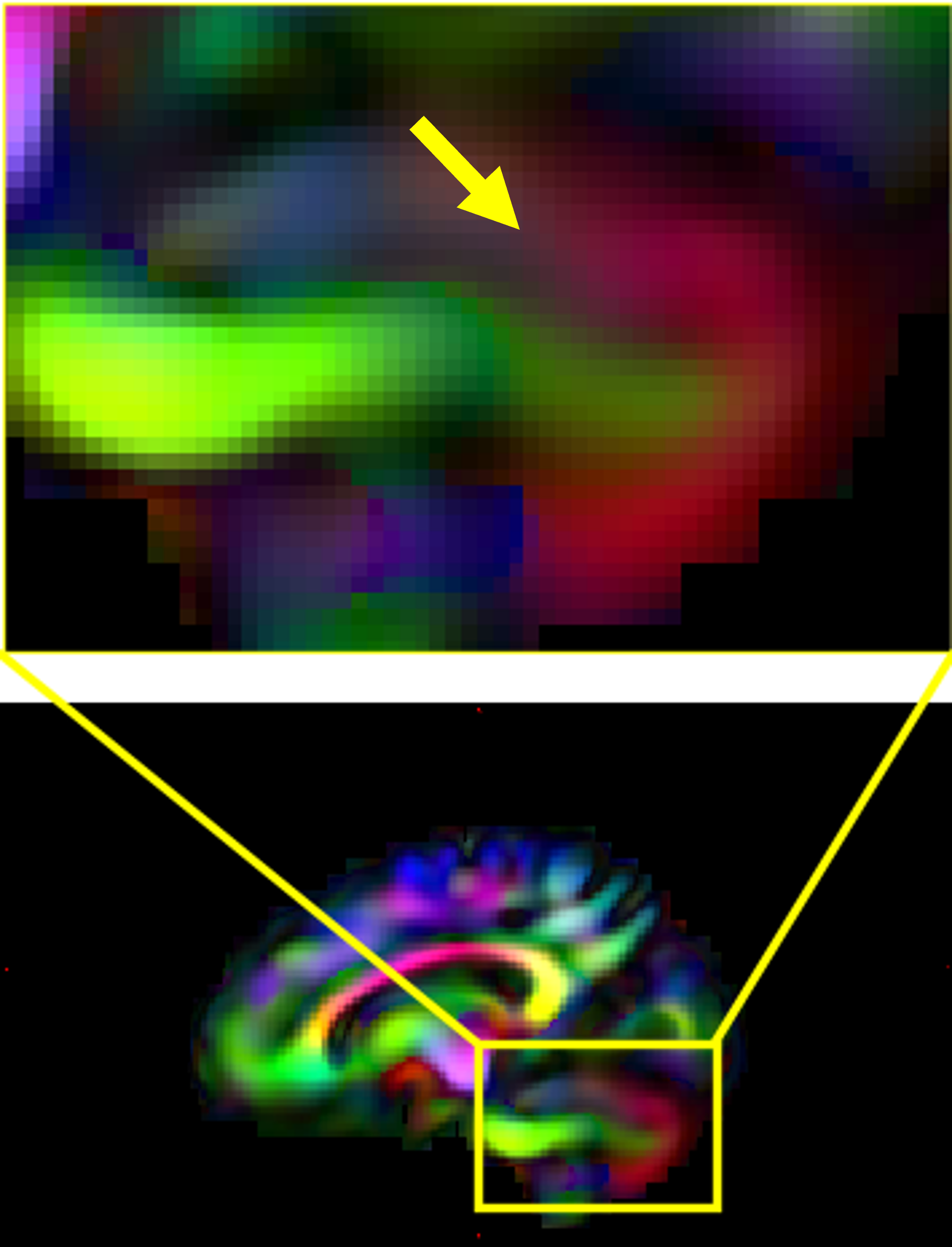}
    \end{minipage}
    \begin{minipage}[c]{0.188\textwidth}
        \centering
        \includegraphics[width=\textwidth]{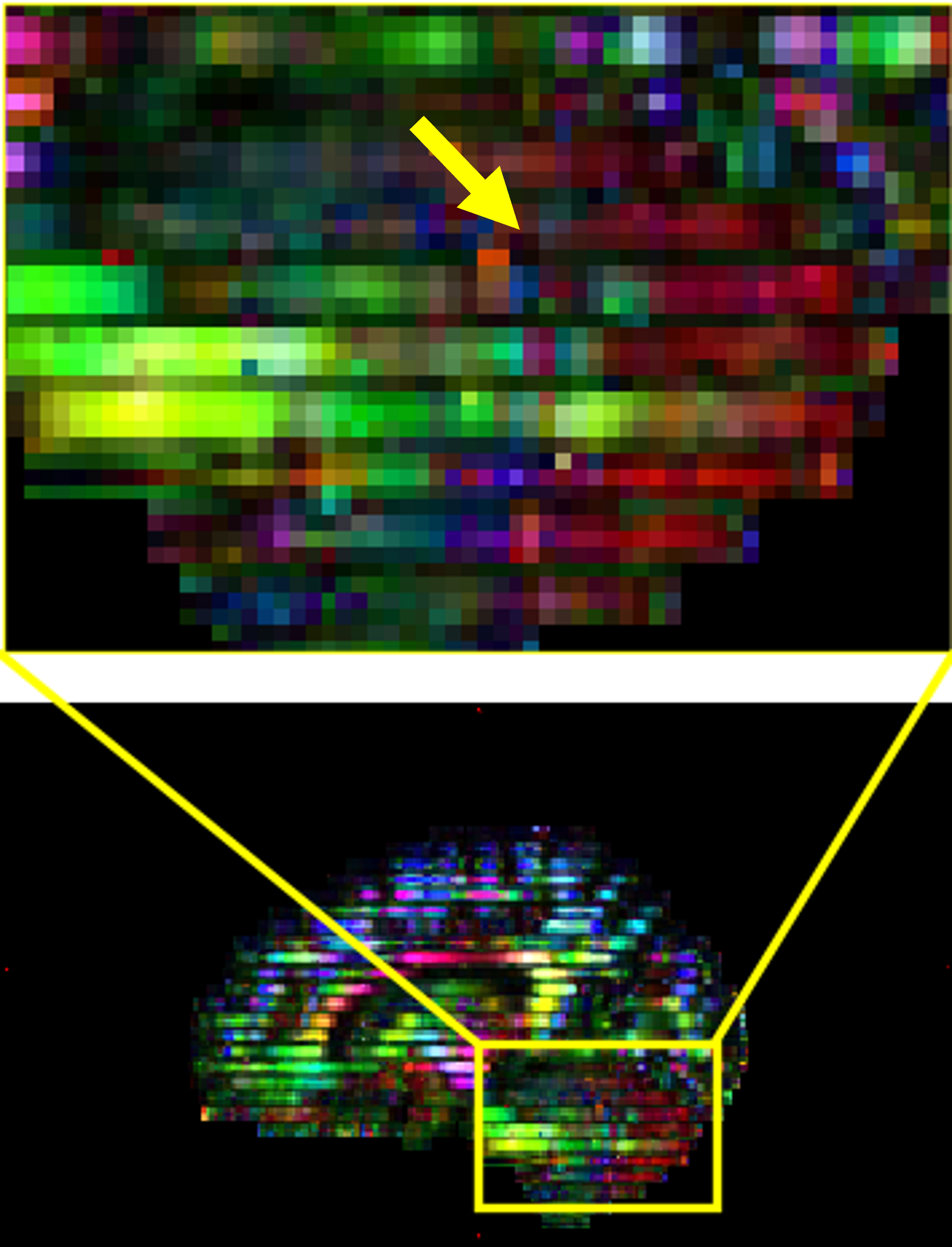}
    \end{minipage}
    \begin{minipage}[c]{0.188\textwidth}
        \centering
        \includegraphics[width=\textwidth]{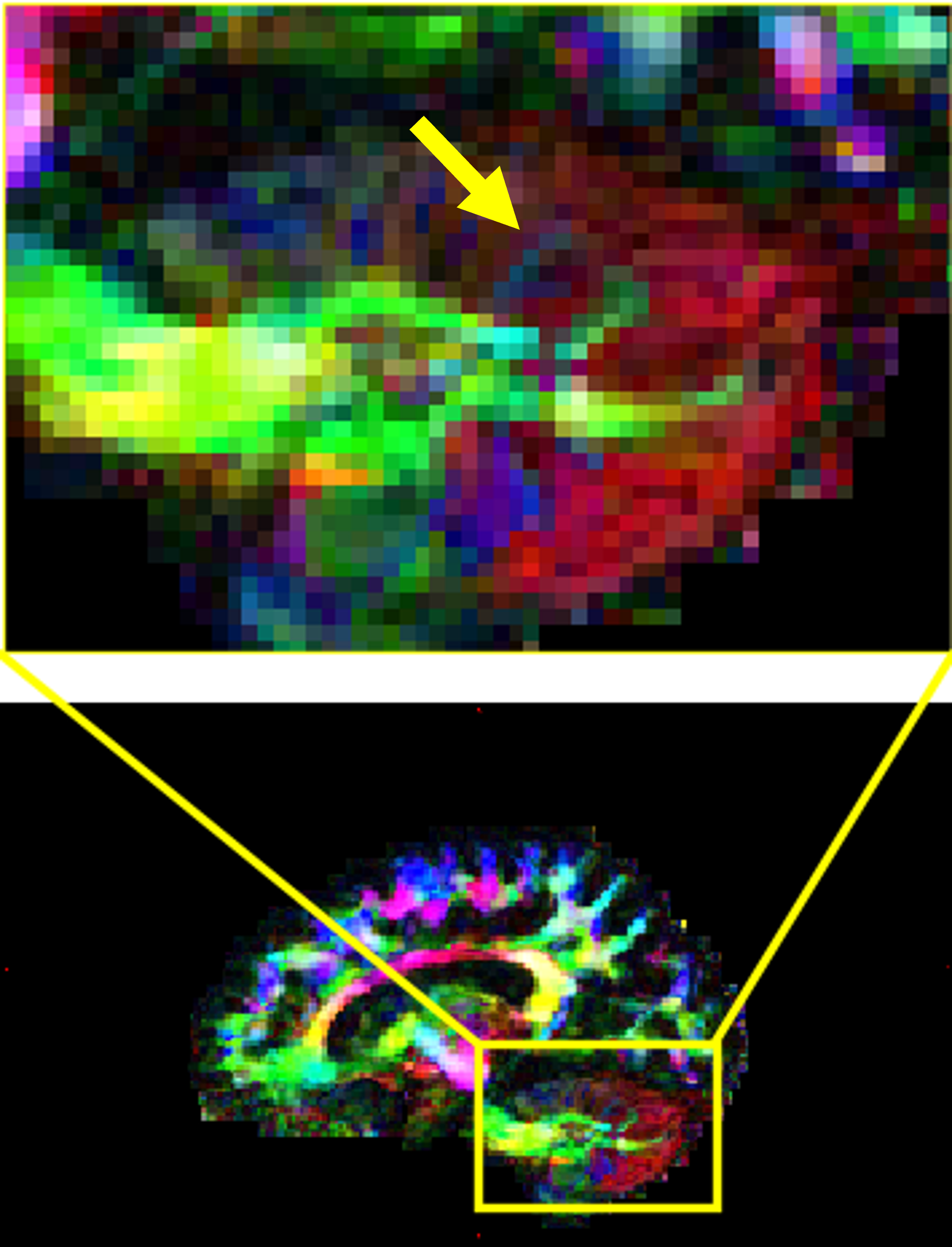}
    \end{minipage}
    \begin{minipage}[c]{0.188\textwidth}
        \centering
        \includegraphics[width=\textwidth]{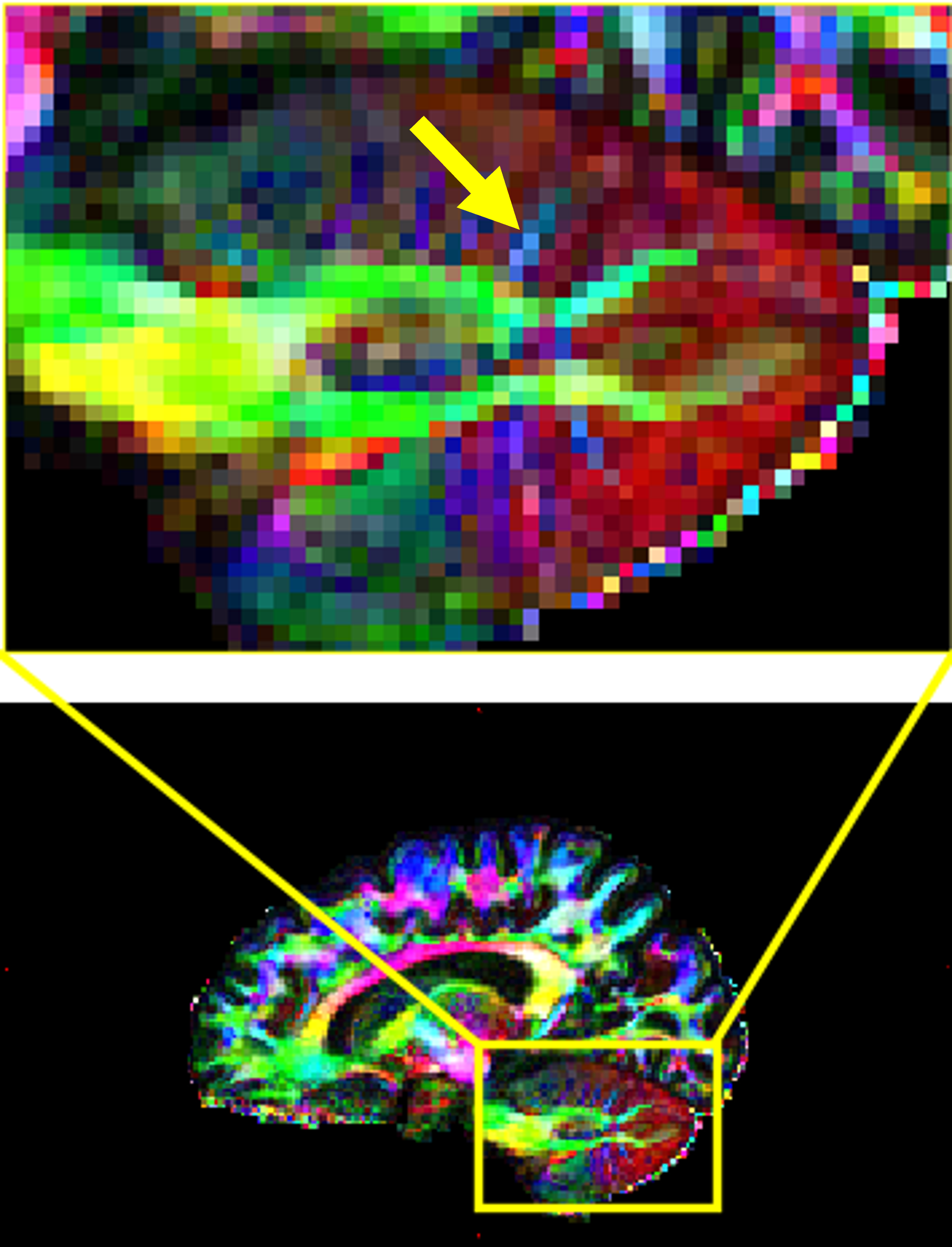}
    \end{minipage}
    
    \par
    
    \begin{minipage}[c]{0.05\textwidth}
        \centering
        \rotatebox[origin=c]{90}{\small fODFs}
    \end{minipage}
    \begin{minipage}[c]{0.188\textwidth}
        \centering
        \includegraphics[width=\textwidth]{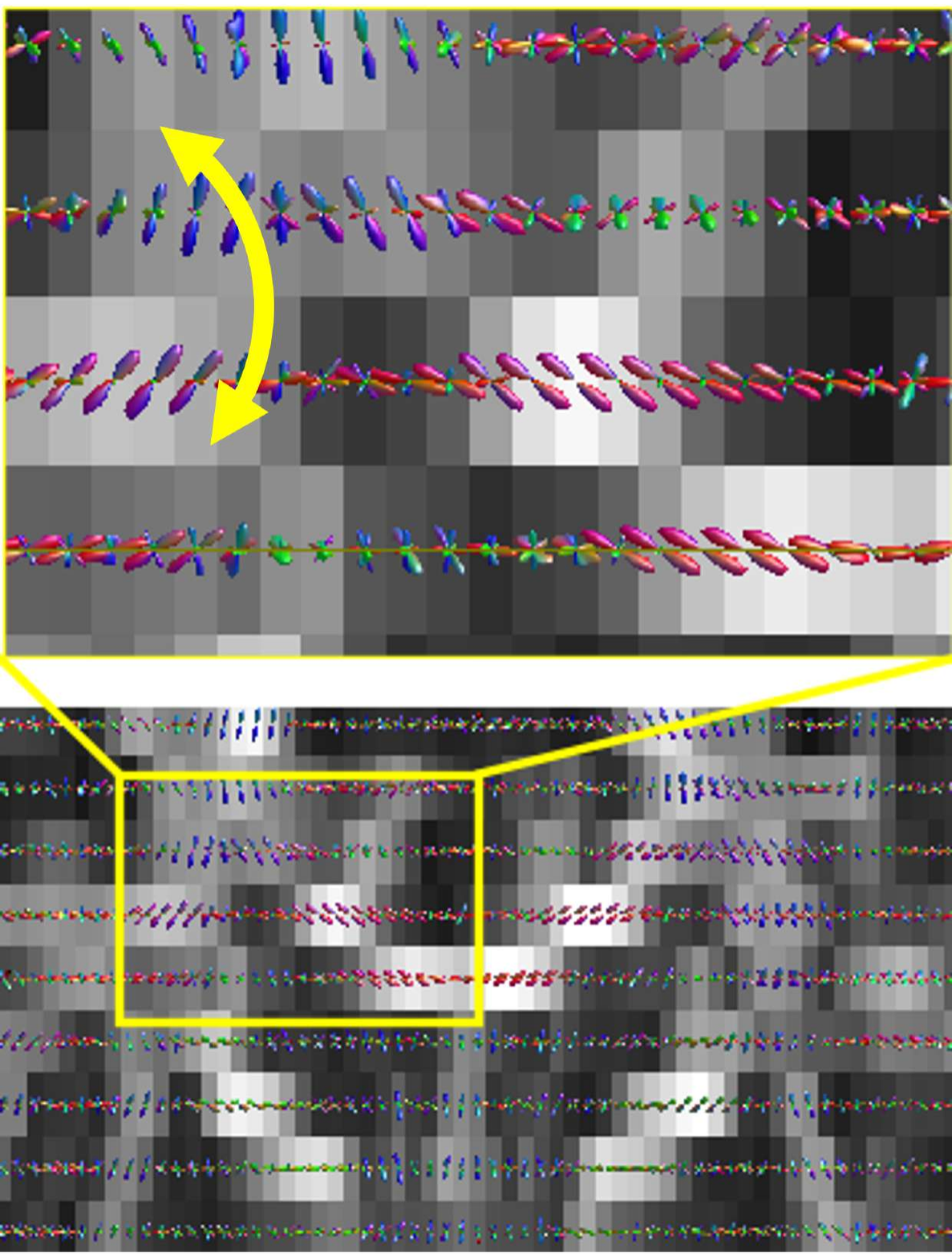}
    \end{minipage}
    \begin{minipage}[c]{0.188\textwidth}
        \centering
        \includegraphics[width=\textwidth]{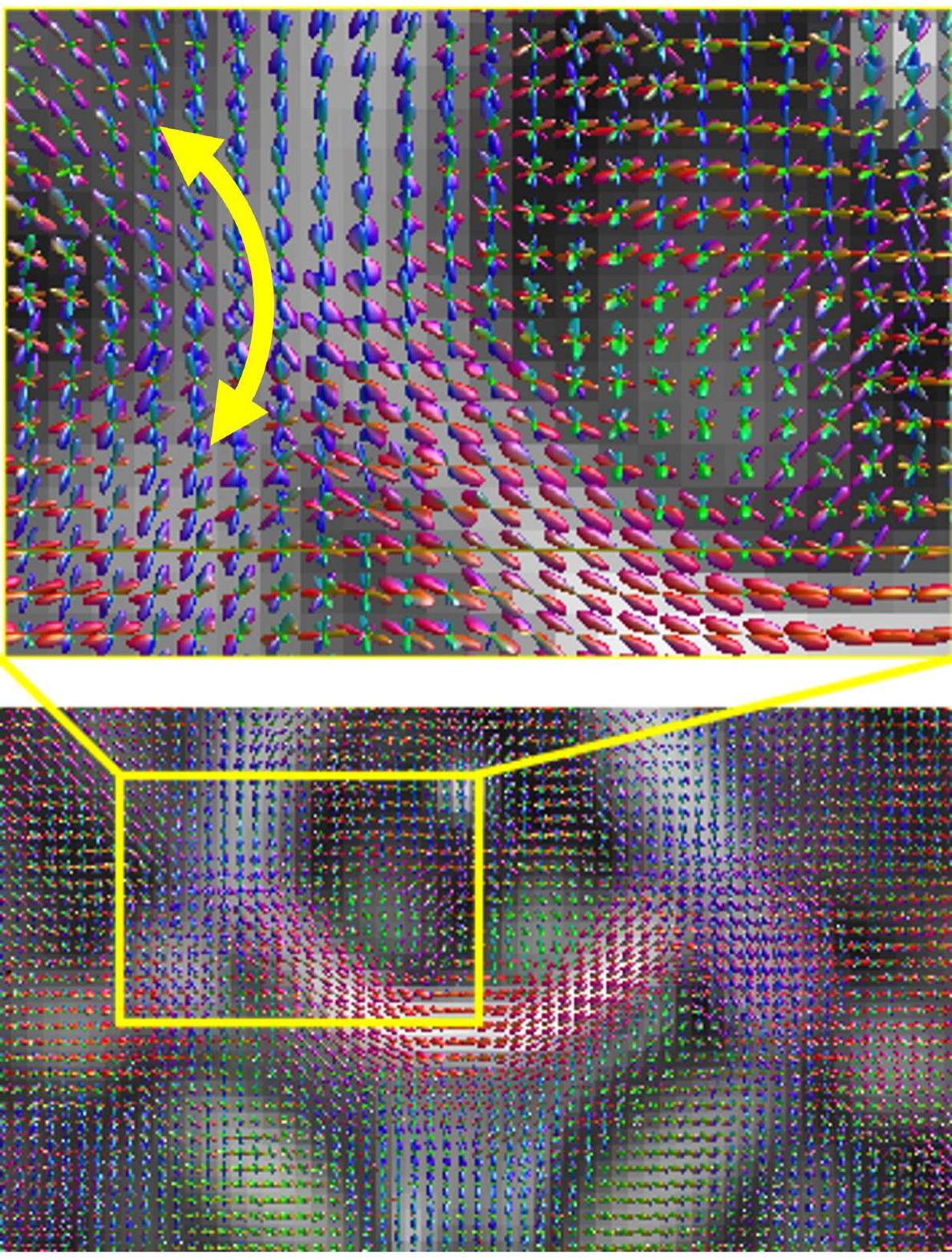}
    \end{minipage}
    \begin{minipage}[c]{0.188\textwidth}
        \centering
        \includegraphics[width=\textwidth]{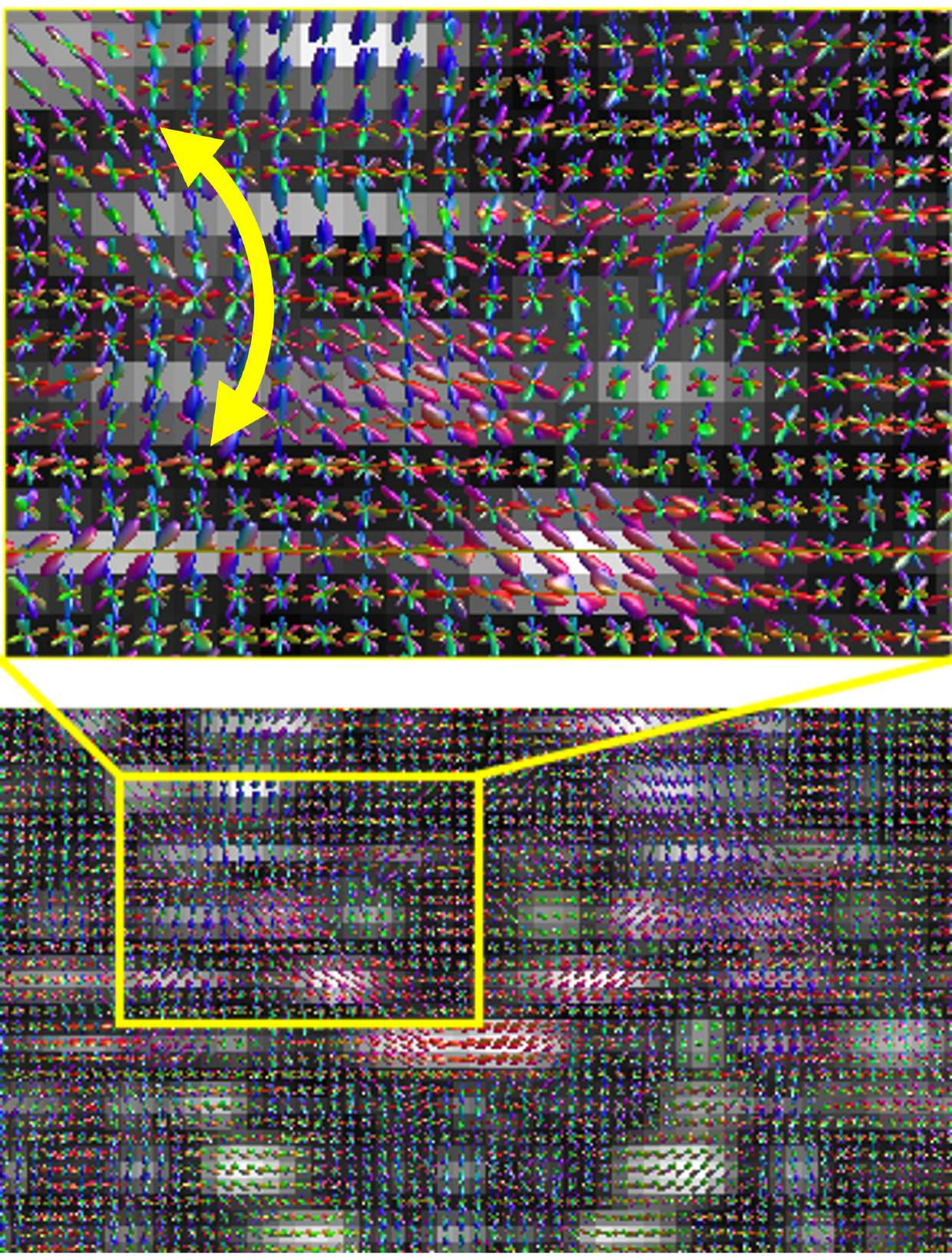}
    \end{minipage}
    \begin{minipage}[c]{0.188\textwidth}
        \centering
        \includegraphics[width=\textwidth]{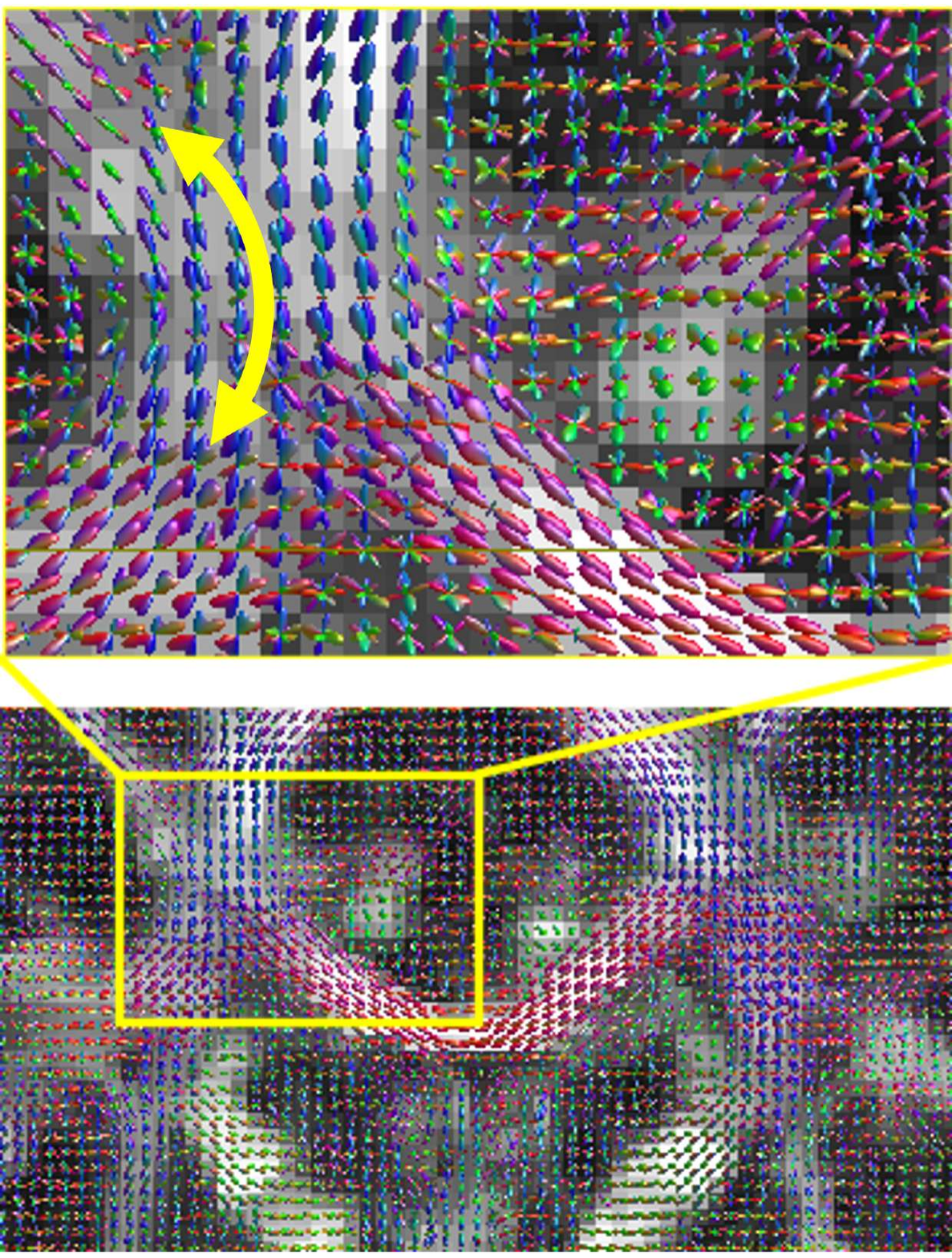}
    \end{minipage}
    \begin{minipage}[c]{0.188\textwidth}
        \centering
        \includegraphics[width=\textwidth]{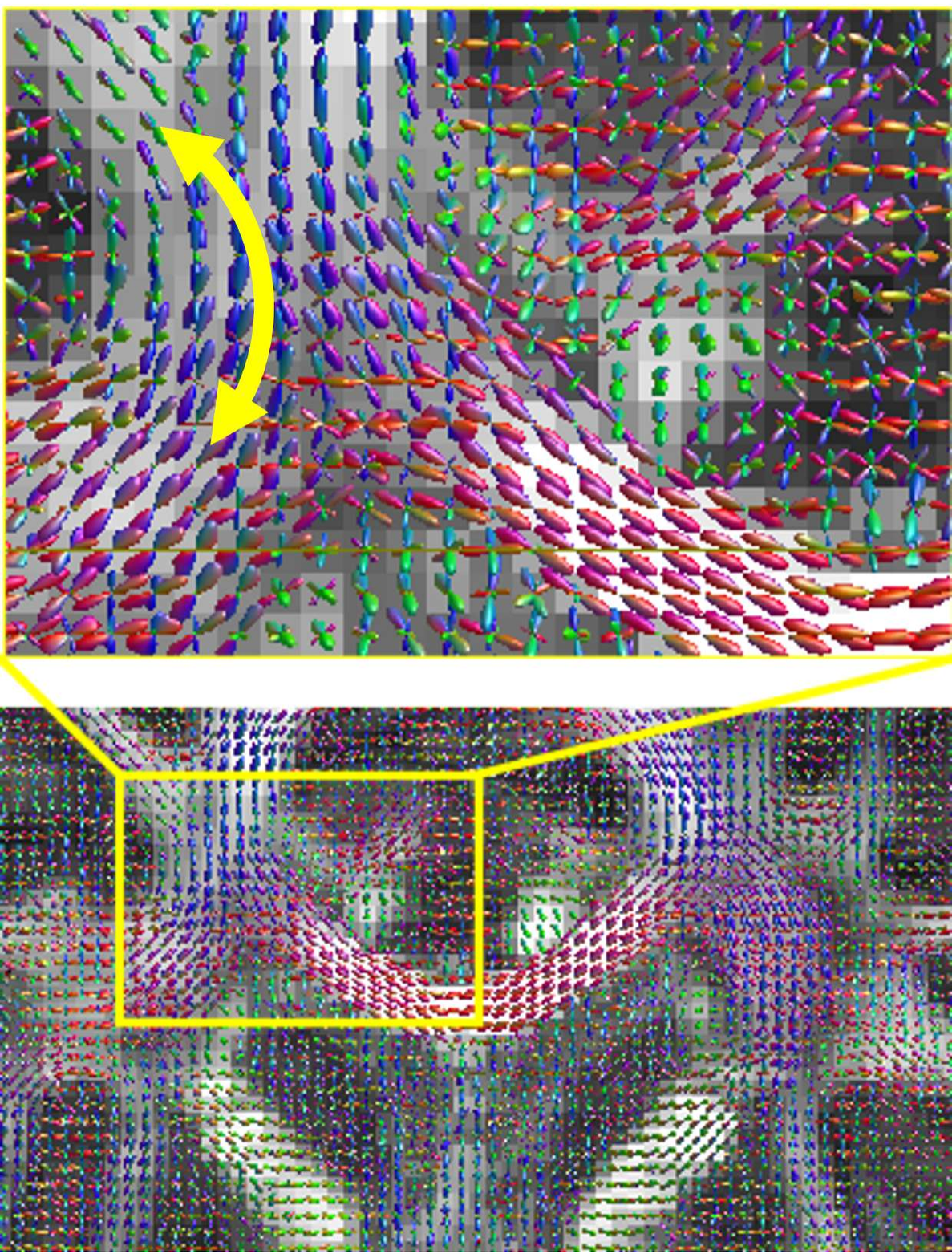}
    \end{minipage}
    
    \par
    
    \begin{minipage}[c]{0.05\textwidth}
        \centering
        \rotatebox[origin=c]{90}{\small Tractography}
    \end{minipage}
    \begin{minipage}[c]{0.188\textwidth}
        \centering
        \sbox0{\includegraphics[width=\textwidth]{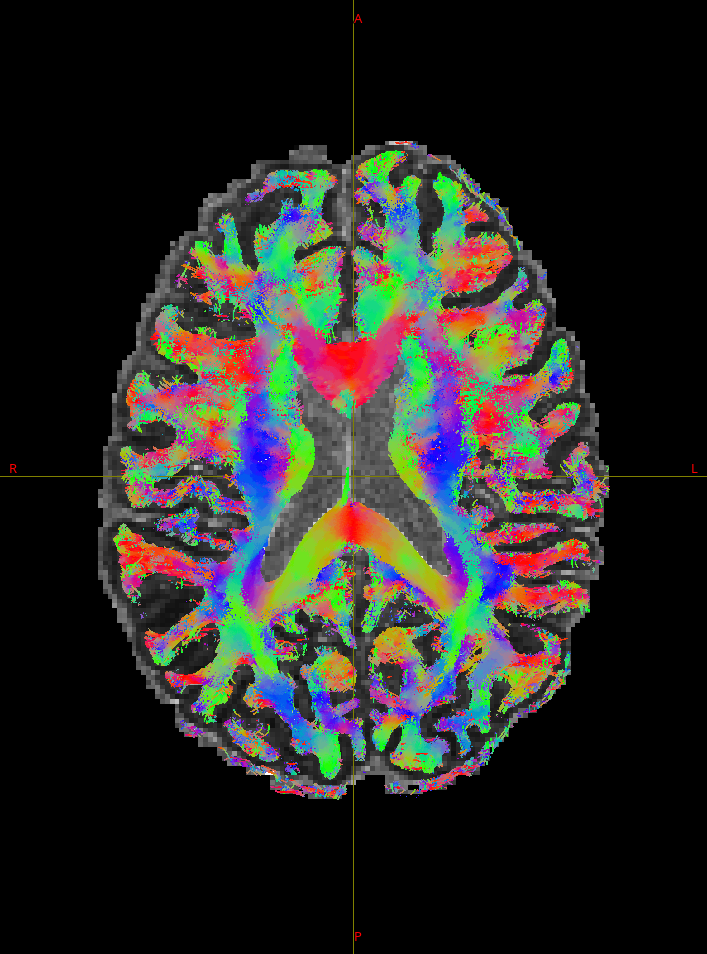}}%
        \setlength{\fboxsep}{0pt}%
        \colorbox{lightgray}{\makebox[\wd0][c]{\parbox[c][\ht0][c]{0.85\wd0}{\centering\footnotesize Slice to be\\recovered\\(not present\\in low\\through-plane\\resolution data)}}}%
    \end{minipage}
    \begin{minipage}[c]{0.188\textwidth}
        \centering
        \includegraphics[width=\textwidth]{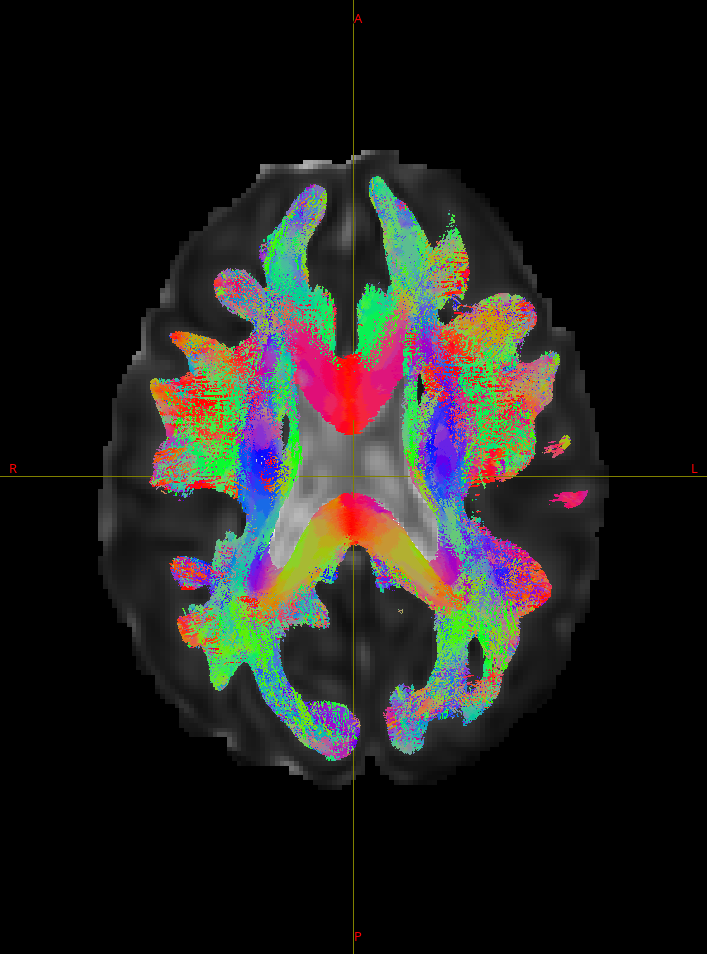}
    \end{minipage}
    \begin{minipage}[c]{0.188\textwidth}
        \centering
        \includegraphics[width=\textwidth]{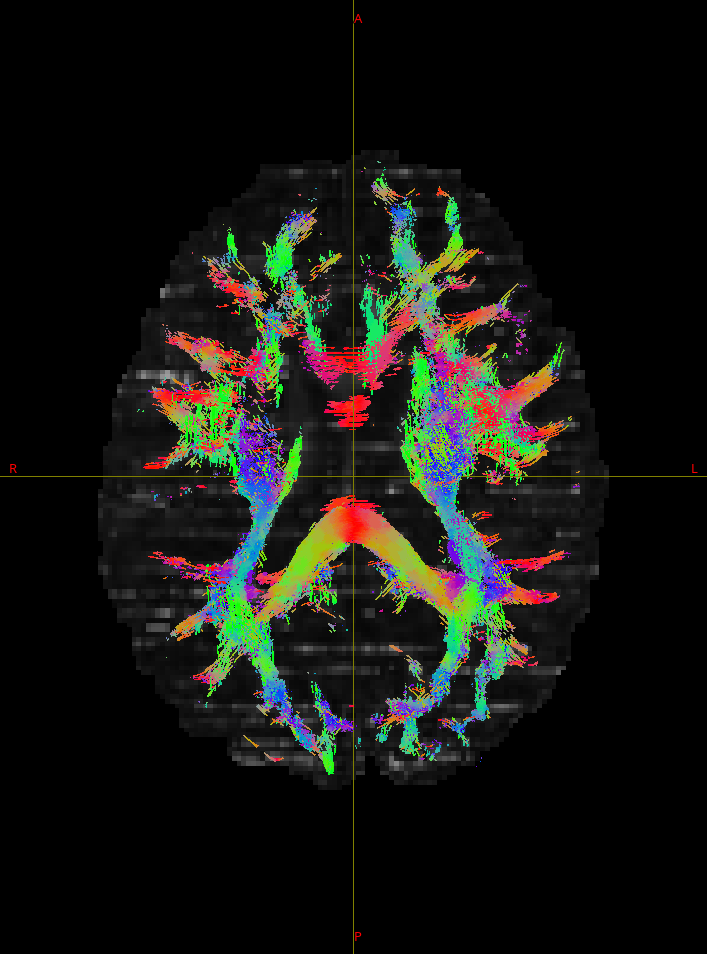}
    \end{minipage}
    \begin{minipage}[c]{0.188\textwidth}
        \centering
        \includegraphics[width=\textwidth]{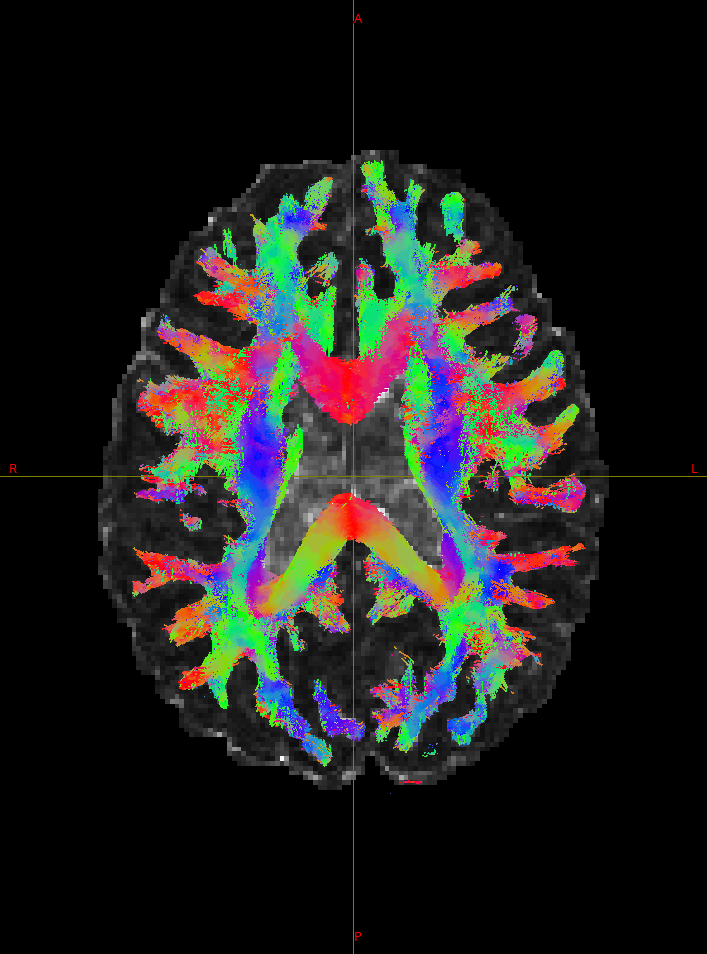}
    \end{minipage}
    \begin{minipage}[c]{0.188\textwidth}
        \centering
        \includegraphics[width=\textwidth]{images/qualitative/Tractography/HighRes0000.png}
    \end{minipage}
    \setlength{\abovecaptionskip}{5pt}
    \caption{\textbf{Reconstruction and runtime.} The figure shows DTI (sagittal, zoomed cerebellum), fODFs overlaid on GFA (coronal, arrow indicates bundles), and tractography (axial) for a missing slice from anisotropic data. Our method recovers fine details in 5 min (plus ${\approx}7$ min initialization), closer to GT than both baselines trained for 1 h.}
    \label{fig:qualitative}
\end{figure*}

 
The top rows of Figure~\ref{fig:qualitative} compare the spatial super-resolution quality using DTI-derived images. After 1 hour of training, NODF produces noticeably blurry reconstructions, while NODF-HashEnc produces sharper reconstructions but introduces blocking artifacts consistent with the hash-grid parameterization. In contrast, our method achieves reconstructions that much more closely match the GT images after only 5 minutes of fine-tuning. The middle rows of Figure~\ref{fig:qualitative} show the fODF fields obtained by running CSD \cite{tournier2008resolving} on the super-resolved dMRI. Echoing the DTI results, we see that our method more closely reconstructs the spatio-angular content in the GT. The bottom row of Figure~\ref{fig:qualitative} shows the tractography results obtained from the reconstructed fODF fields. While NODF is able to reconstruct most of the major pathways, it misses some tracts near the white-gray matter interface, likely due to the oversmoothing effect observed in the DTI images. NODF-HashEnc is able to reconstruct more tracts near the white-gray interface compared to NODF, but suffers from an overall reduction in streamlines, likely due to early termination from the blocking artifacts. Our method produces tracts that much more closely resemble those obtained from the GT, with higher streamline density and better coverage near the white-gray matter interface.

\subsection{Generalization to Clinical Data}
\begin{figure}[!t]
    \centering

    \hspace{0.04\textwidth}
    \begin{minipage}[t]{0.187\textwidth}\centering
        \small\textbf{Raw Signal}
    \end{minipage}%
    \hspace{0.005\textwidth}%
    \begin{minipage}[t]{0.187\textwidth}\centering
        \small\textbf{Raw DTI}
    \end{minipage}%
    \hspace{0.005\textwidth}%
    \begin{minipage}[t]{0.187\textwidth}\centering
        \small\textbf{Interpolation}
    \end{minipage}%
    \hspace{0.005\textwidth}%
    \begin{minipage}[t]{0.187\textwidth}\centering
        \small\textbf{Registration}
    \end{minipage}%
    \hspace{0.005\textwidth}%
    \begin{minipage}[t]{0.187\textwidth}\centering
        \small\textbf{Proposed}
    \end{minipage}

    \vspace{0.005\textwidth}

    \makebox[0.04\textwidth]{\raisebox{4em}[0pt][0pt]{\rotatebox[origin=c]{90}{\small Axial}}}%
    \begin{minipage}[t]{0.187\textwidth}
        \centering
        \includegraphics[width=\textwidth]{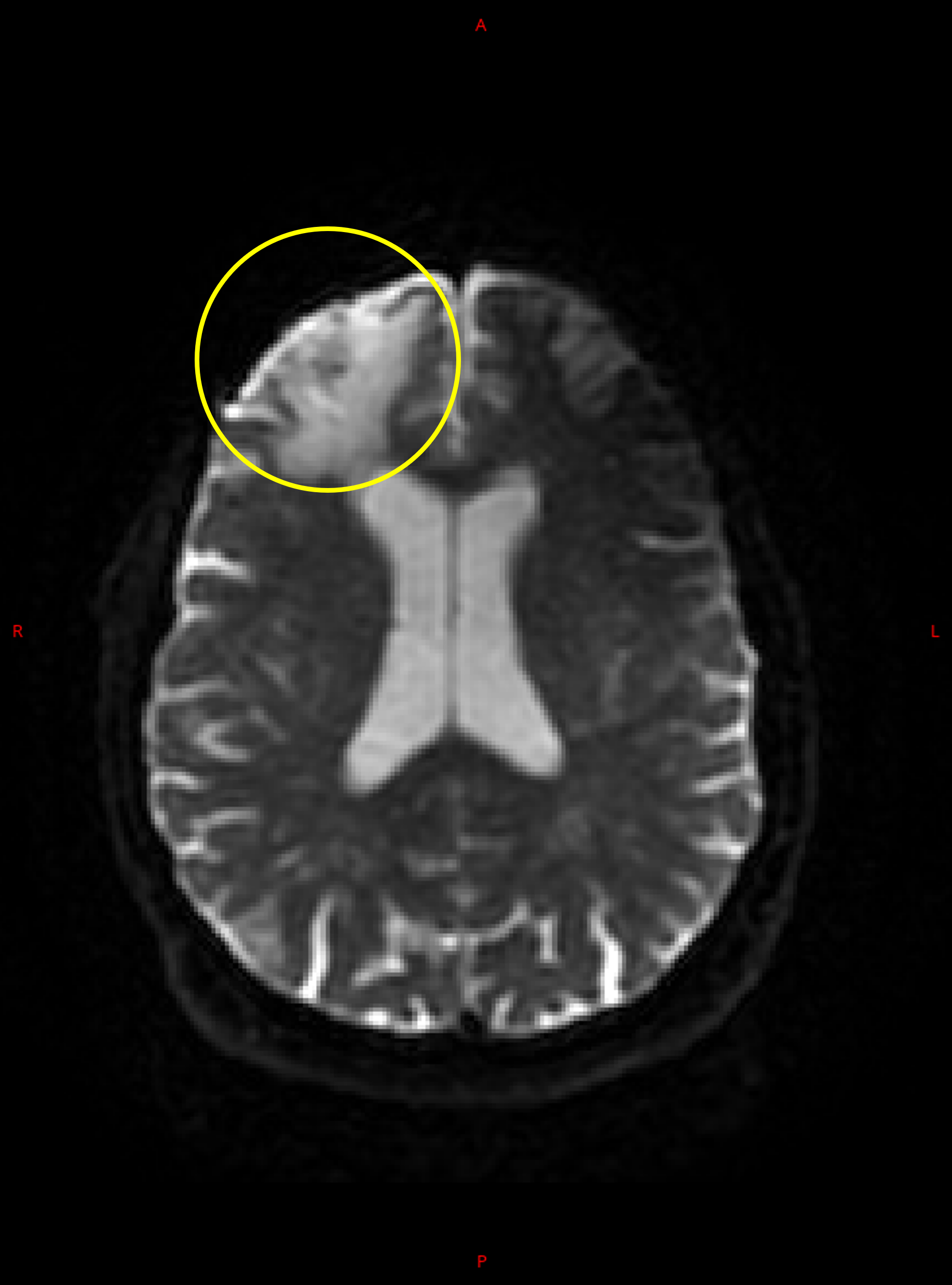}
    \end{minipage}%
    \hspace{0.005\textwidth}%
    \begin{minipage}[t]{0.187\textwidth}
        \centering
        \includegraphics[width=\textwidth]{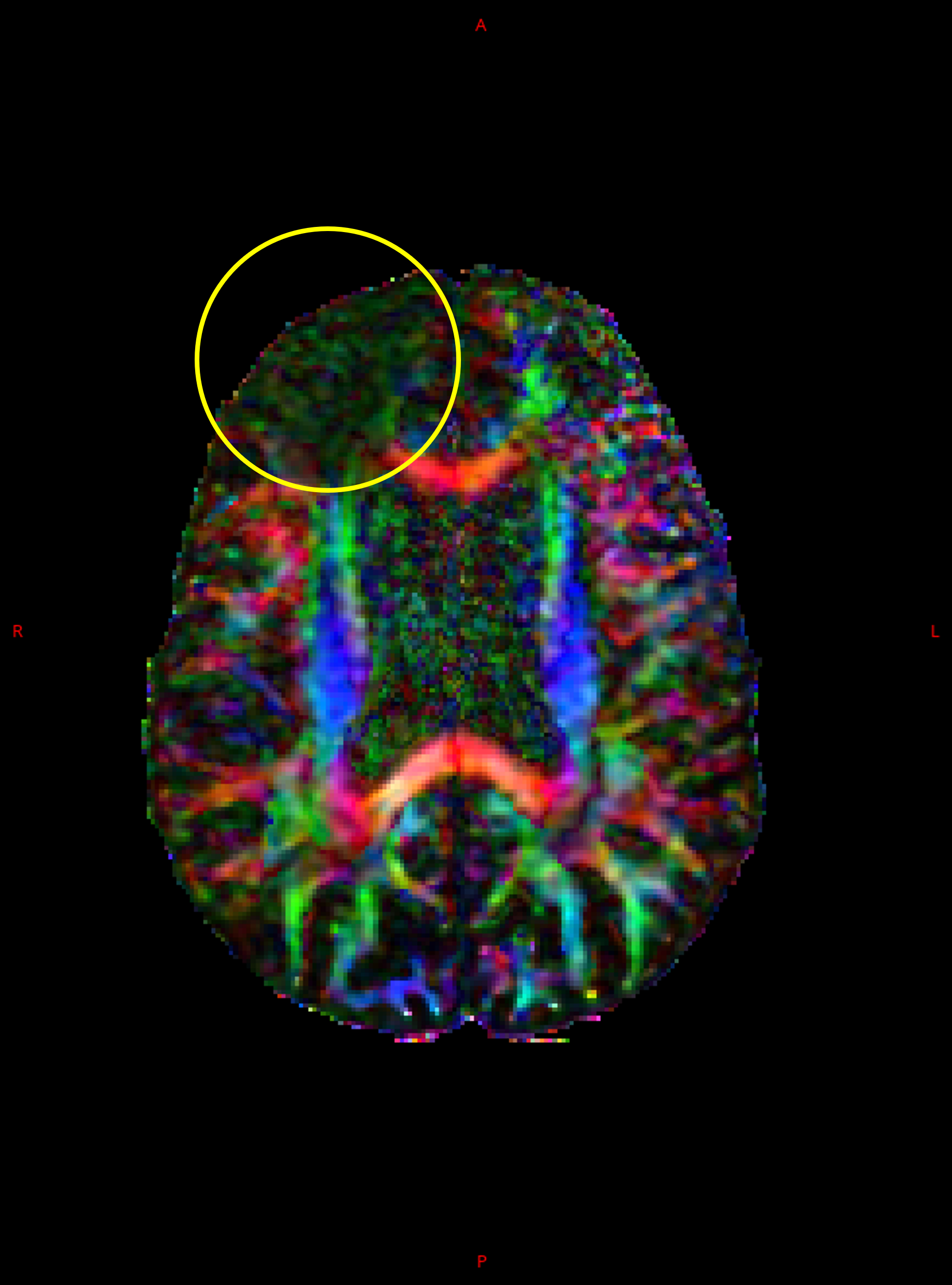}
    \end{minipage}%
    \hspace{0.005\textwidth}%
    \begin{minipage}[t]{0.187\textwidth}
        \centering
        \includegraphics[width=\textwidth]{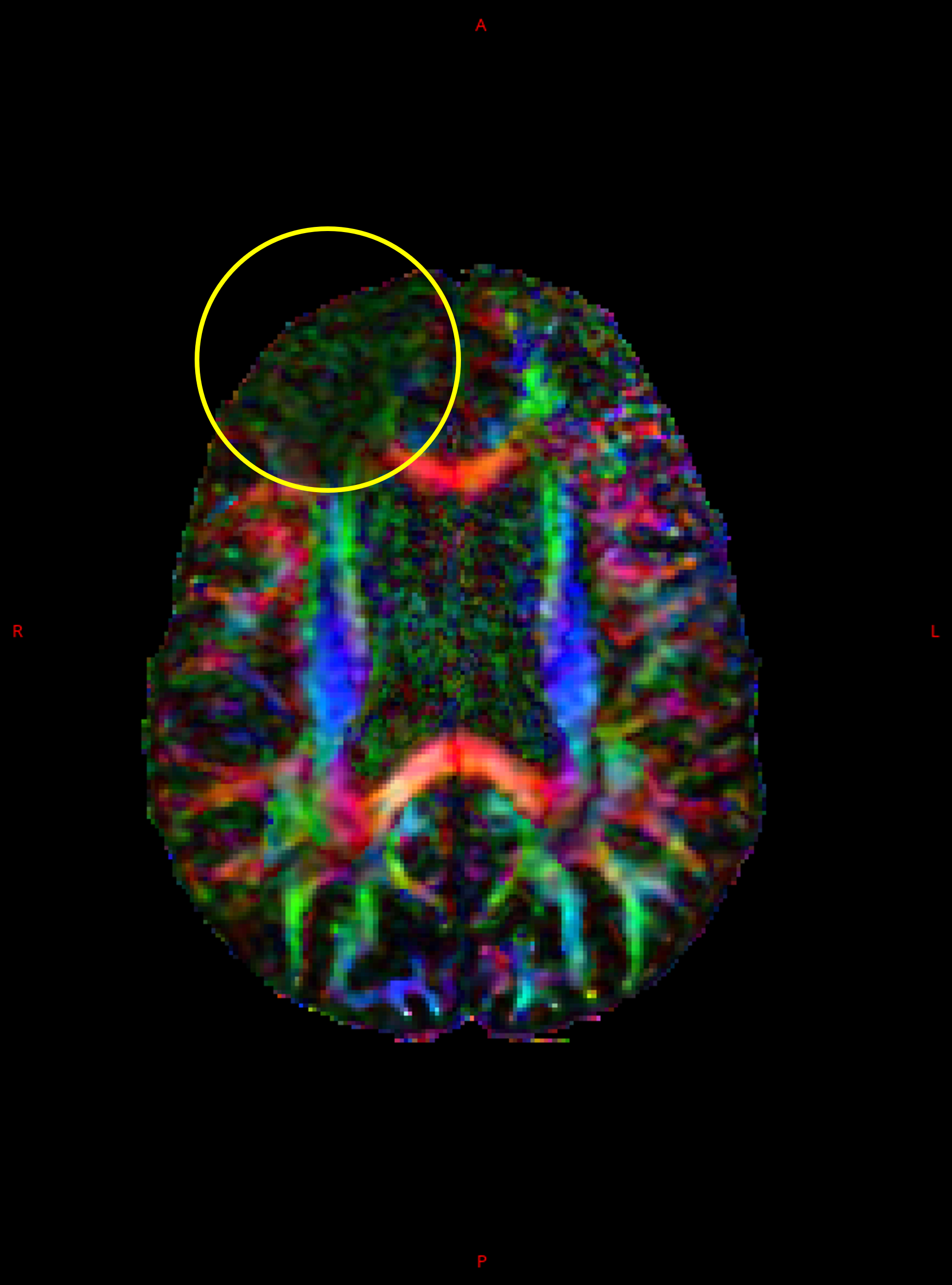}
    \end{minipage}%
    \hspace{0.005\textwidth}%
    \begin{minipage}[t]{0.187\textwidth}
        \centering
        \includegraphics[width=\textwidth]{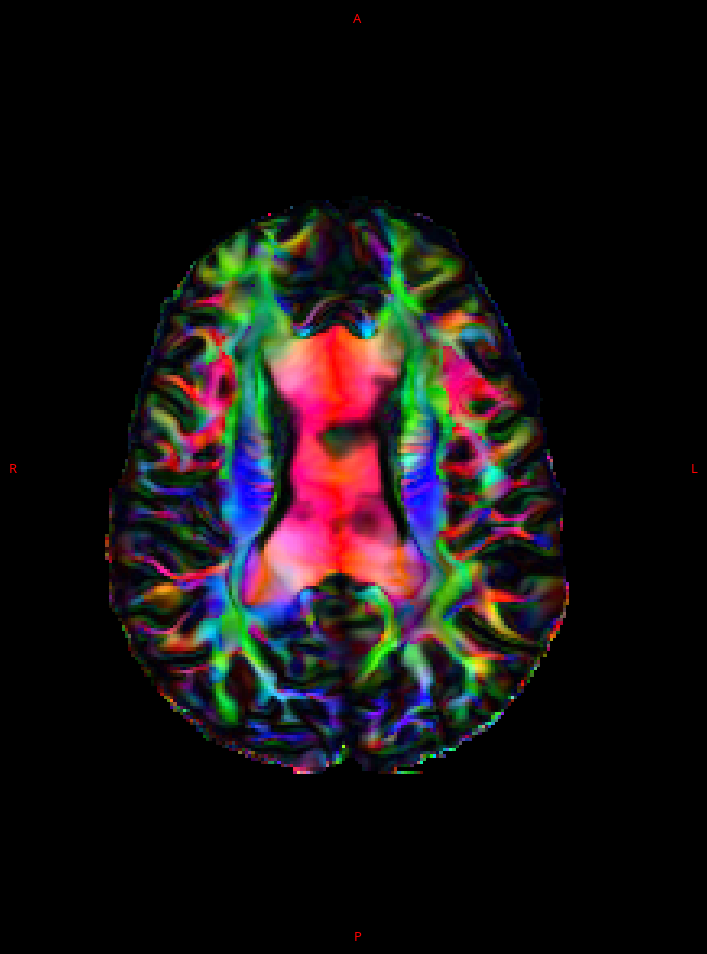}
    \end{minipage}%
    \hspace{0.005\textwidth}%
    \begin{minipage}[t]{0.187\textwidth}
        \centering
        \includegraphics[width=\textwidth]{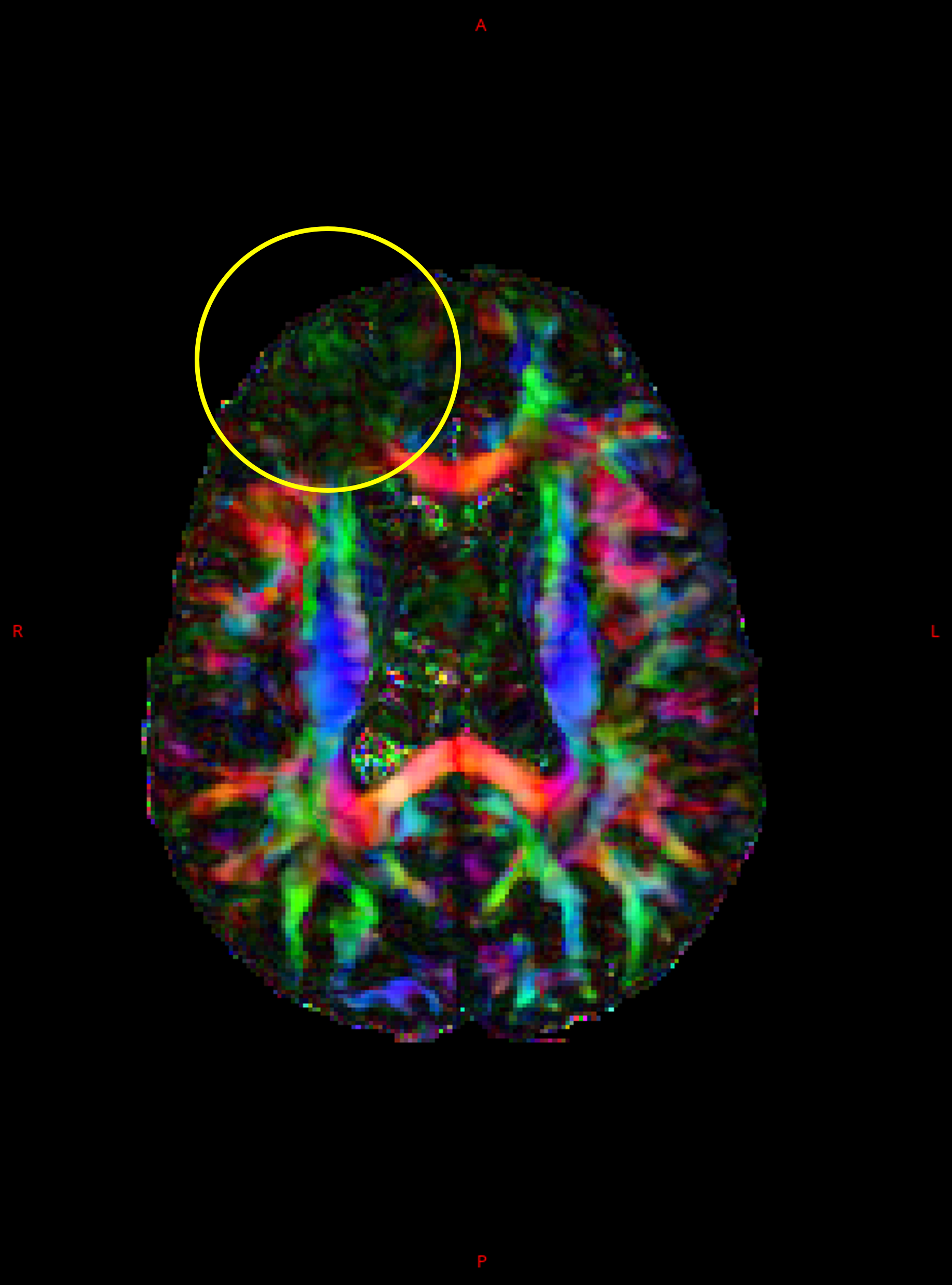}
    \end{minipage}

    \vspace{0.005\textwidth}

    \makebox[0.04\textwidth]{\raisebox{4em}[0pt][0pt]{\rotatebox[origin=c]{90}{\small Sagittal}}}%
    \begin{minipage}[t]{0.187\textwidth}
        \centering
        \includegraphics[width=\textwidth]{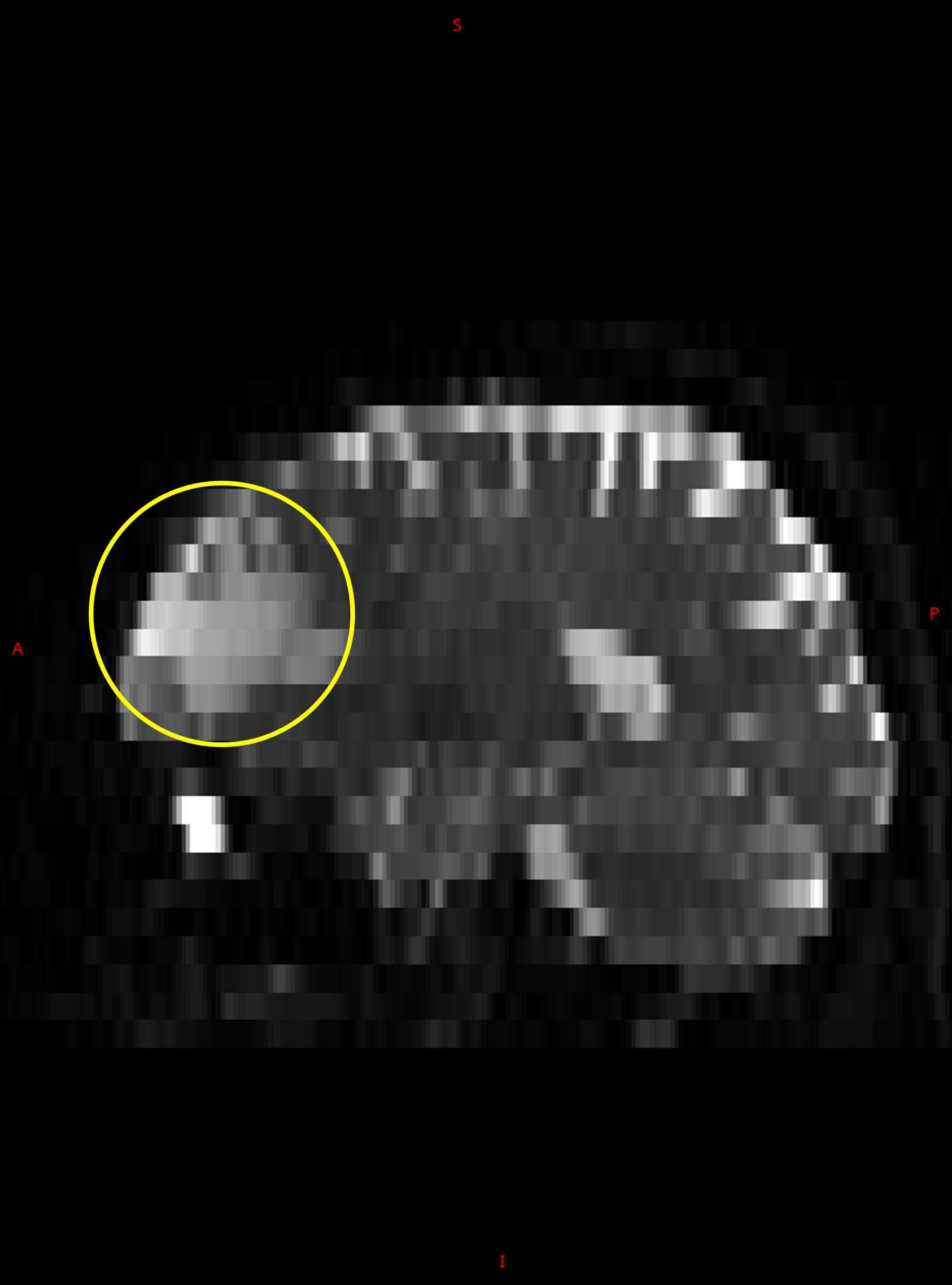}
    \end{minipage}%
    \hspace{0.005\textwidth}%
    \begin{minipage}[t]{0.187\textwidth}
        \centering
        \includegraphics[width=\textwidth]{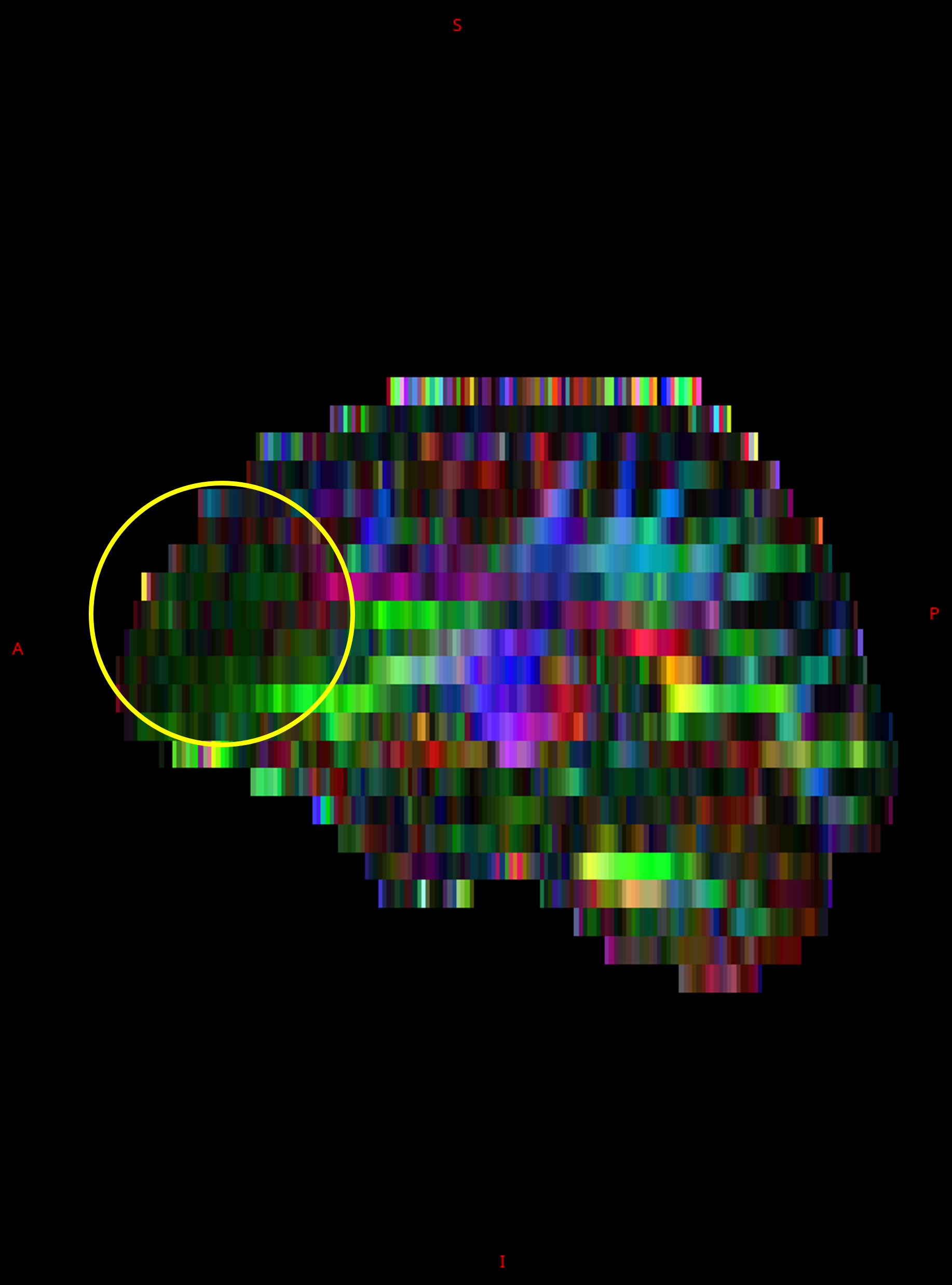}
    \end{minipage}%
    \hspace{0.005\textwidth}%
    \begin{minipage}[t]{0.187\textwidth}
        \centering
        \includegraphics[width=\textwidth]{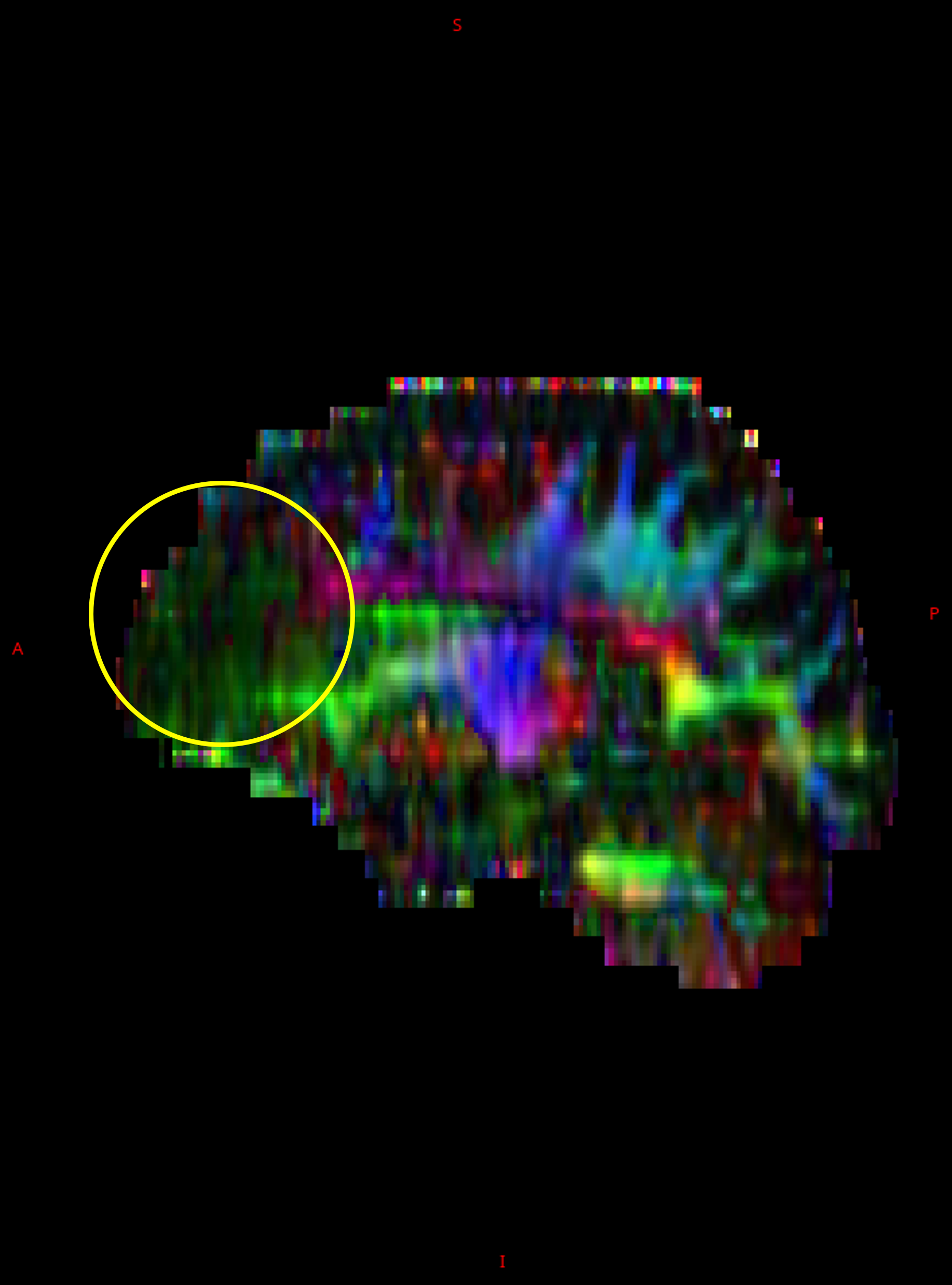}
    \end{minipage}%
    \hspace{0.005\textwidth}%
    \begin{minipage}[t]{0.187\textwidth}
        \centering
        \includegraphics[width=\textwidth]{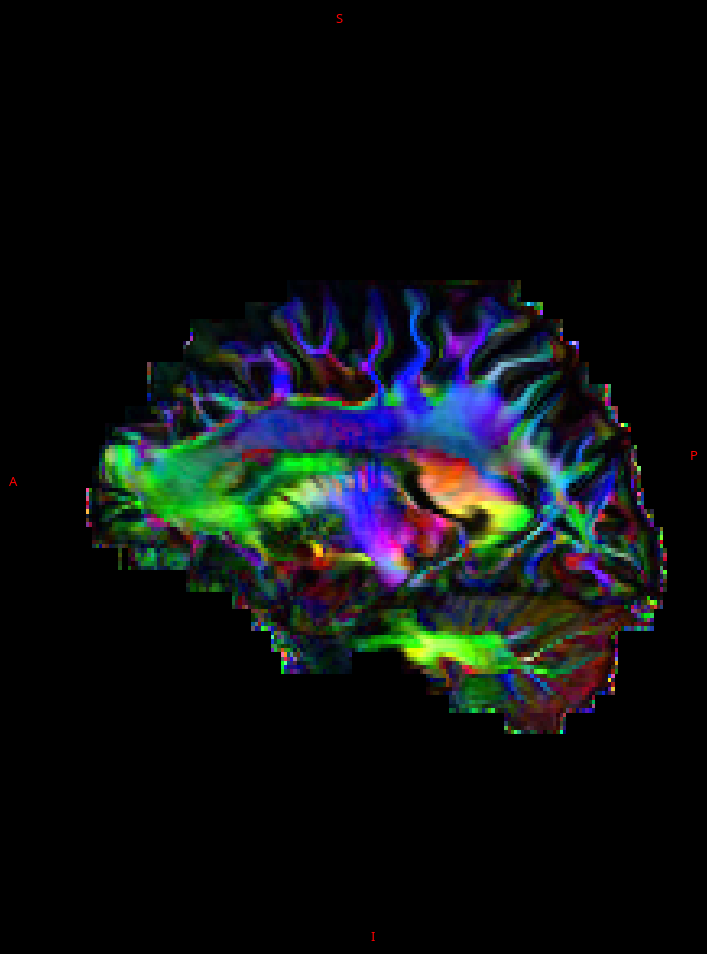}
    \end{minipage}%
    \hspace{0.005\textwidth}%
    \begin{minipage}[t]{0.187\textwidth}
        \centering
        \includegraphics[width=\textwidth]{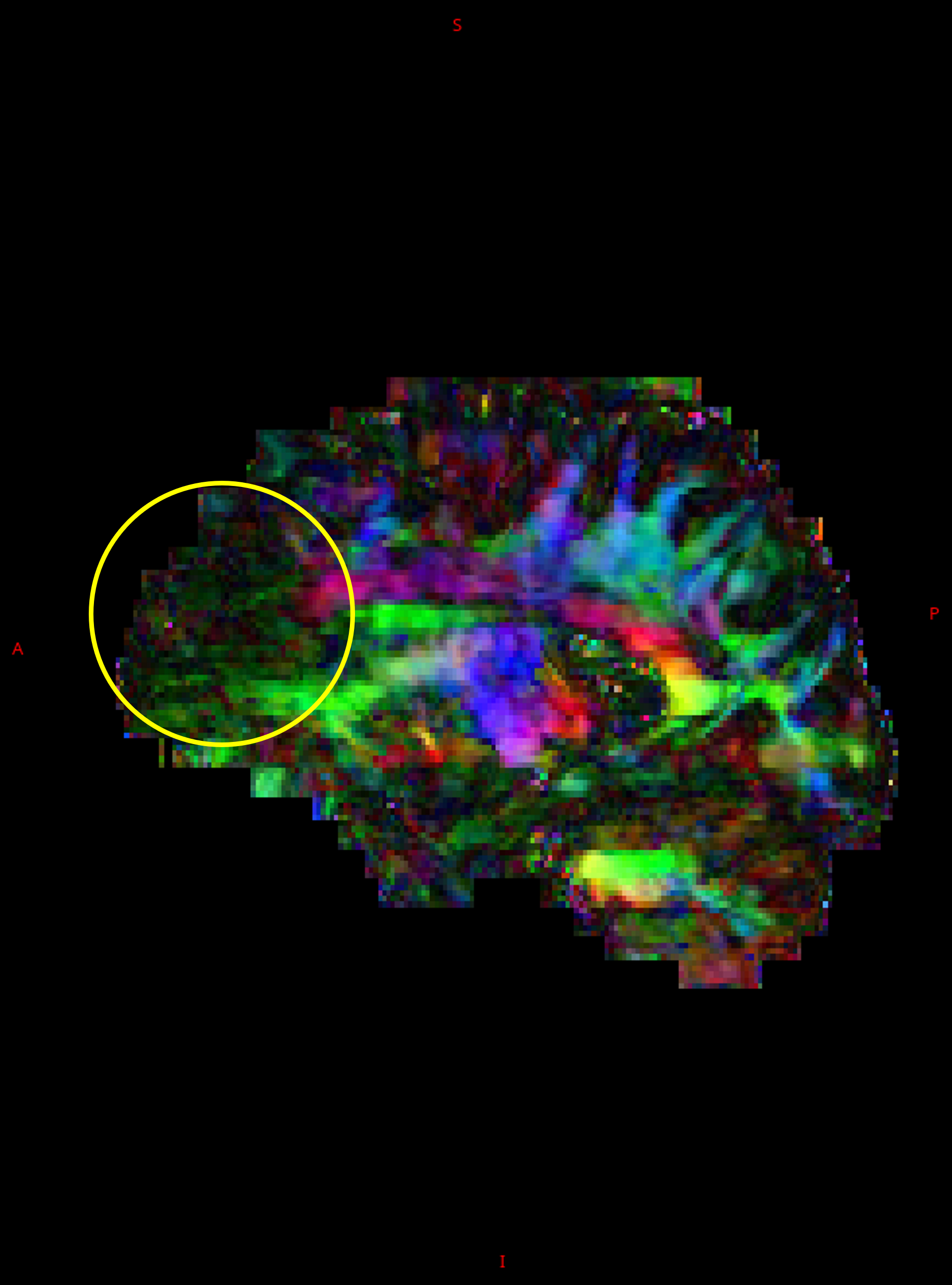}
    \end{minipage}

    \par\vspace{0.005\textwidth}

    \makebox[0.04\textwidth]{\raisebox{4em}[0pt][0pt]{\rotatebox[origin=c]{90}{\small Coronal}}}%
    \begin{minipage}[t]{0.187\textwidth}
        \centering
        \includegraphics[width=\textwidth]{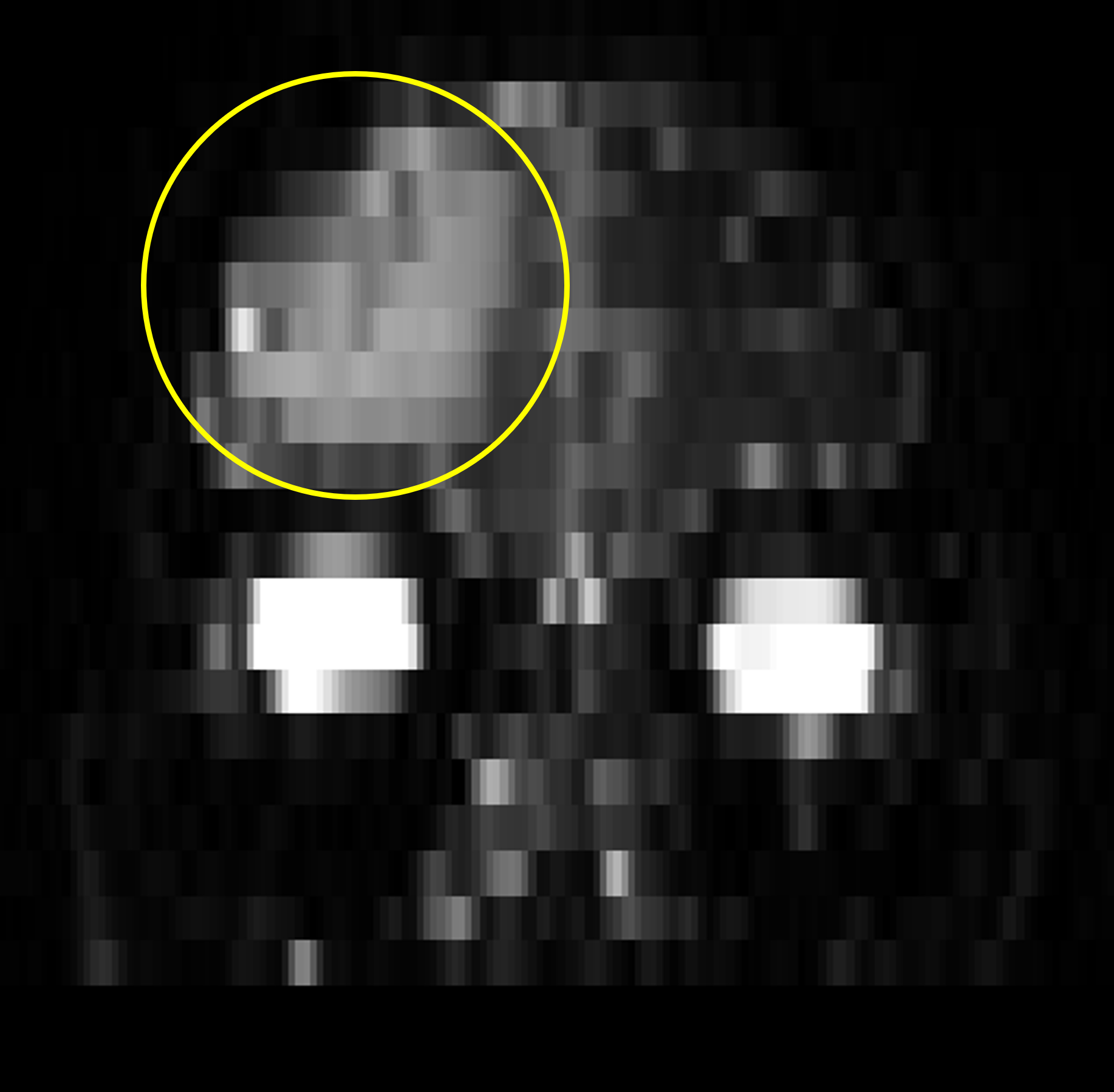}
    \end{minipage}%
    \hspace{0.005\textwidth}%
    \begin{minipage}[t]{0.187\textwidth}
        \centering
        \includegraphics[width=\textwidth]{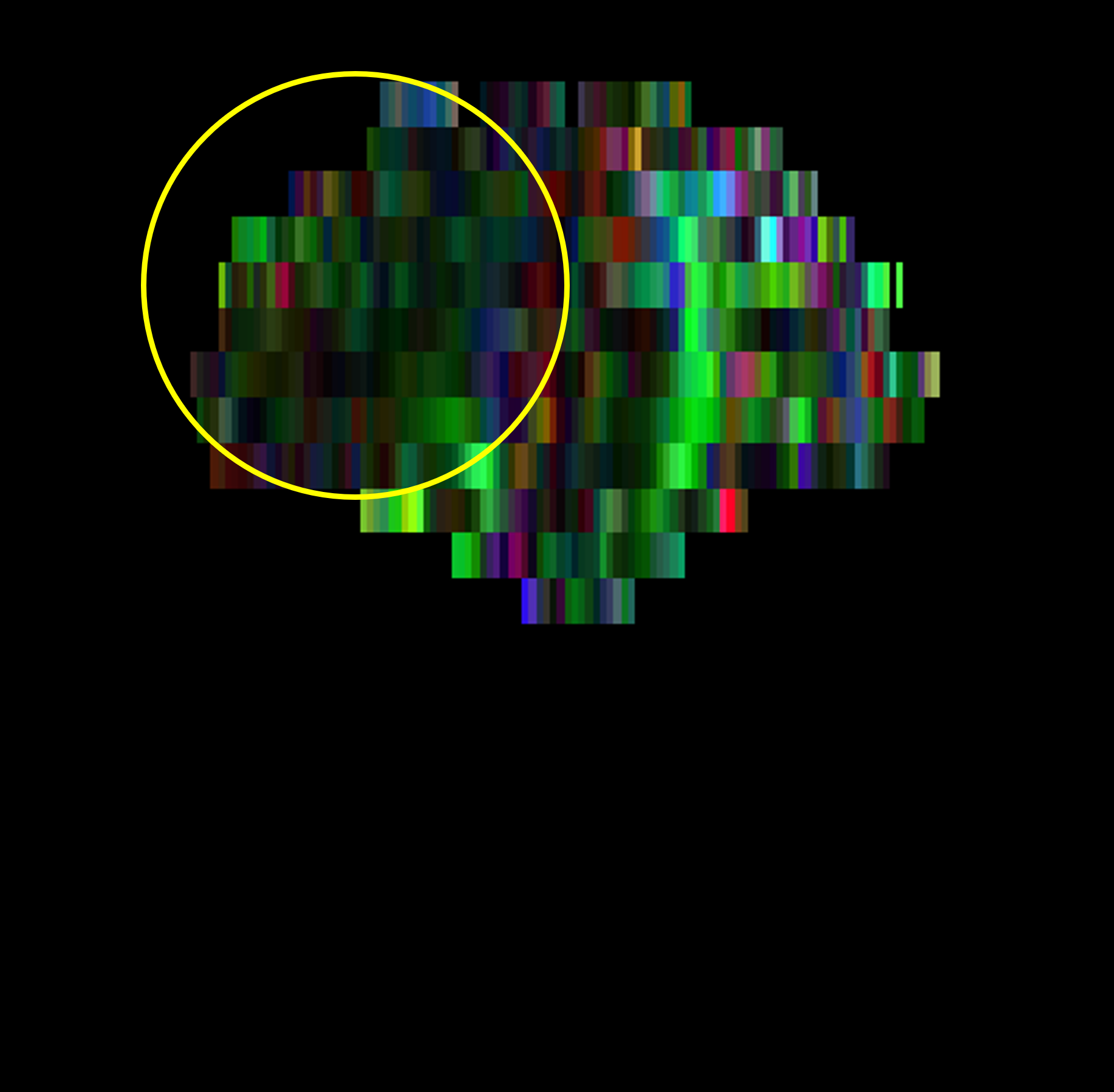}
    \end{minipage}%
    \hspace{0.005\textwidth}%
    \begin{minipage}[t]{0.187\textwidth}
        \centering
        \includegraphics[width=\textwidth]{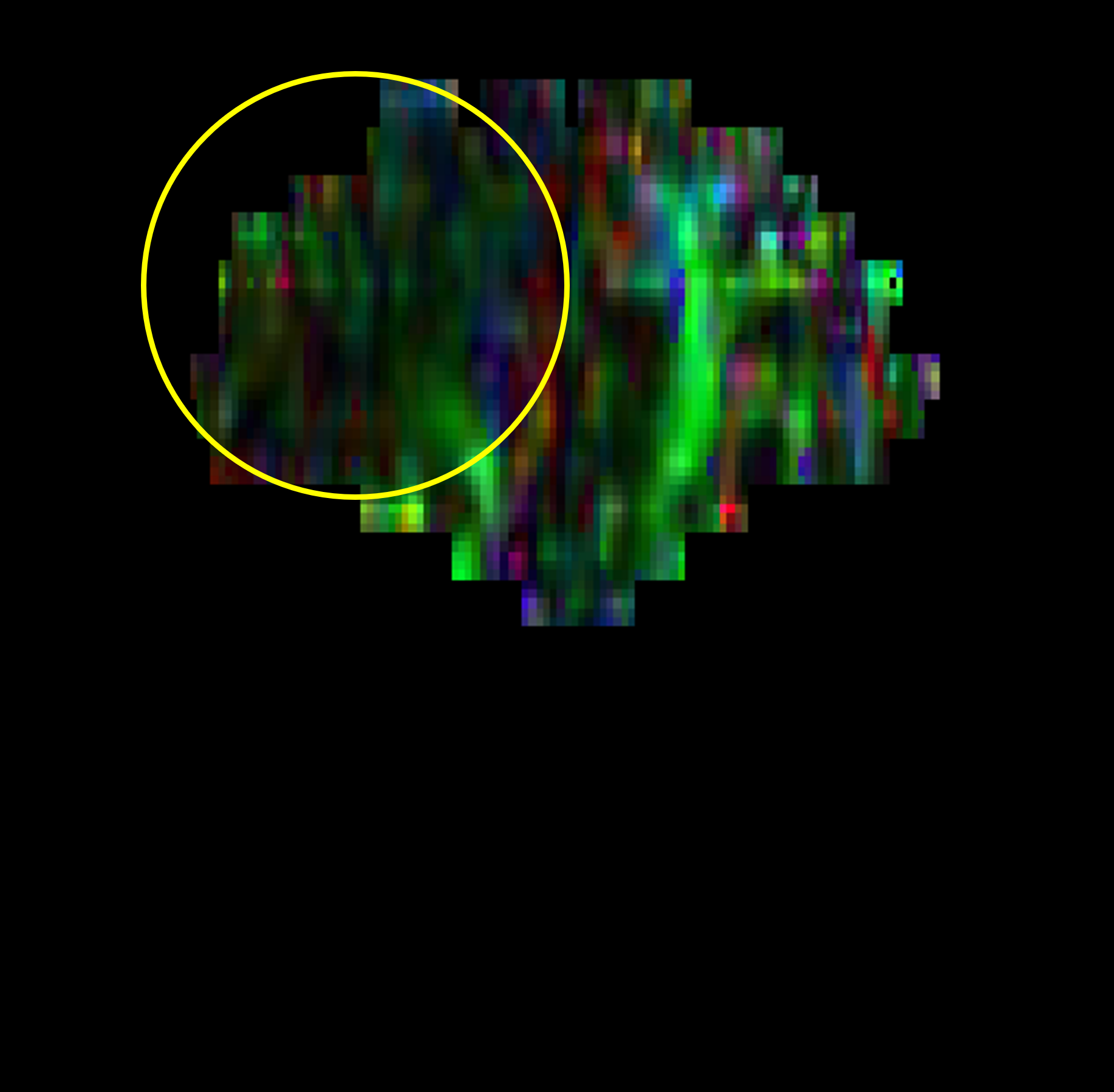}
    \end{minipage}%
    \hspace{0.005\textwidth}%
    \begin{minipage}[t]{0.187\textwidth}
        \centering
        \includegraphics[width=\textwidth]{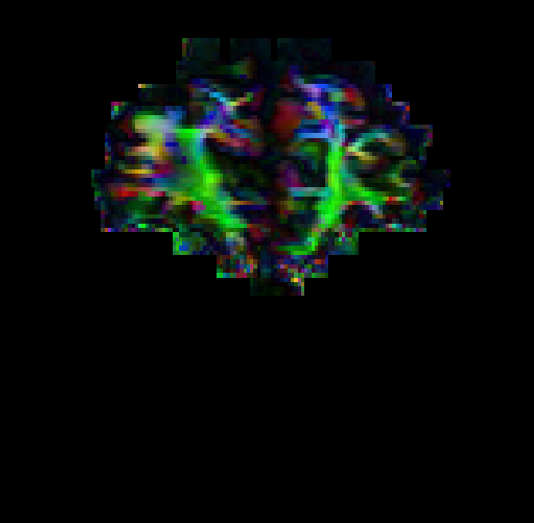}
    \end{minipage}%
    \hspace{0.005\textwidth}%
    \begin{minipage}[t]{0.187\textwidth}
        \centering
        \includegraphics[width=\textwidth]{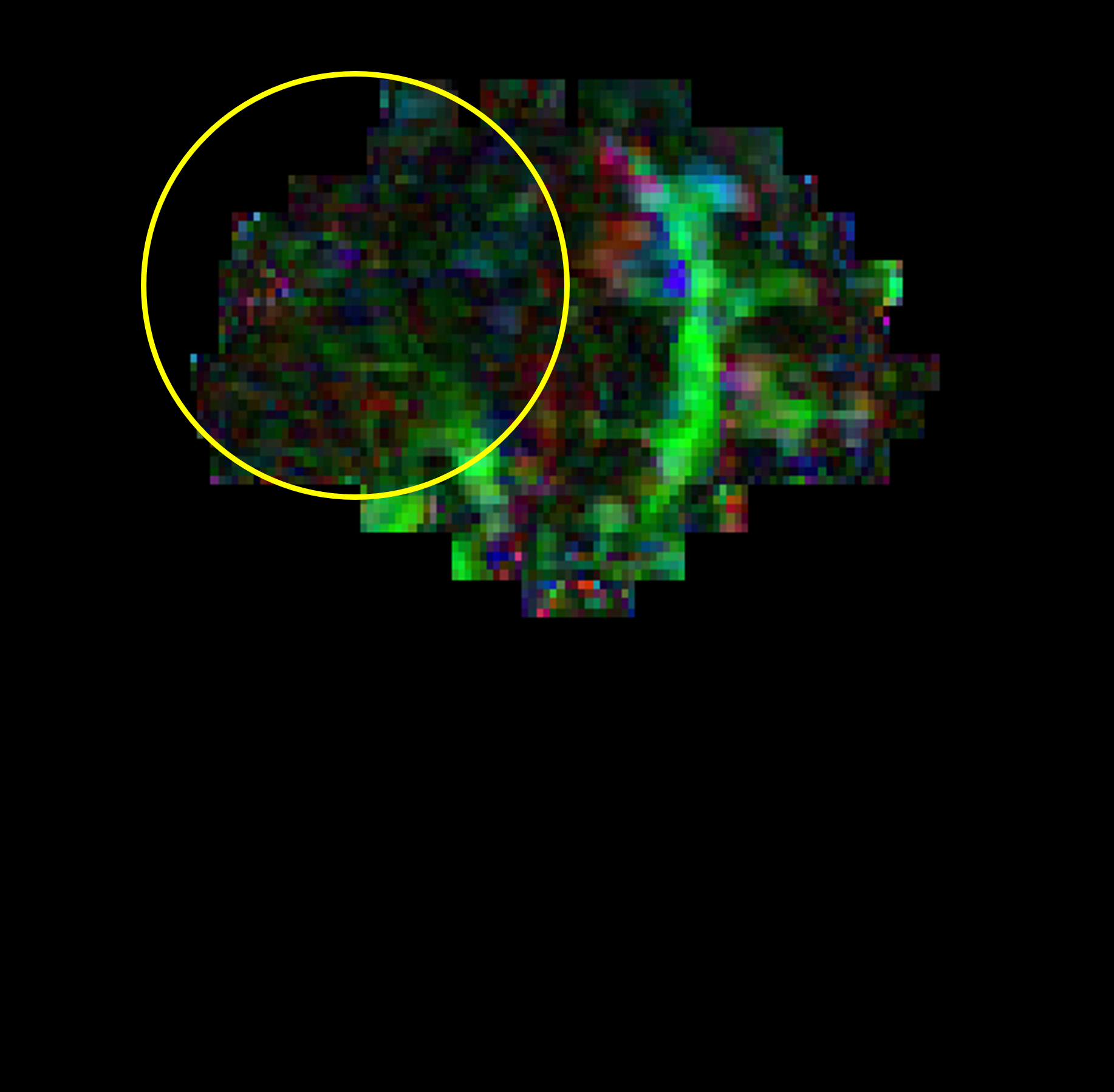}
    \end{minipage}

    \setlength{\abovecaptionskip}{5pt}
    \caption{Qualitative evaluation on a real clinical dMRI acquisition. Yellow circles mark the lesion in the right anterior region of the brain. The third column shows trilinear interpolation. The lesion is absent in the fourth column (Stages~1--2 of Section~\ref{sec:methodology}), where the healthy template prior has not yet been adapted to subject-specific anatomy via fine-tuning. Non-brain regions (e.g., eyes) have been masked in columns 2--5. Appropriate informed consent was obtained for this acquisition.}
    \label{fig:lesion}
\end{figure}

To assess robustness beyond the HCP training distribution, we apply our method to a real clinical dMRI acquisition at $0.9375 \times 0.9375 \times 6$\,mm, for which appropriate informed consent was obtained. The clinical data were processed with \emph{luigi-pnlpipe}~\cite{luigipnlpipe}, a publicly available pipeline. As isotropic ground-truth acquisitions are unavailable for real clinical scans, this evaluation is necessarily qualitative.
\par
Compared to the HCP experiments ($1.25 \times 1.25 \times 5$\,mm input with $4\times$ through-plane super-resolution), this case presents a non-integer $4.8 \times$ through-plane upsampling factor, higher in-plane resolution, and realistic acquisition noise. The super-resolution target is set to $0.9375 \times 0.9375 \times 1.25$\,mm, with the HCP template’s resolution in the through-plane direction and the subject’s higher in-plane resolution. We do not include the template in this comparison because anatomical differences with the clinical subject make a direct visual comparison uninformative. The axial slice shown in the fifth column of Figure~\ref{fig:lesion} corresponds approximately to the same anatomical level as the low-resolution input, with minor positional differences arising from the shift to the finer $1.25$\,mm through-plane resampling grid.
\par
Figure~\ref{fig:lesion} shows DTI color maps in three anatomical views. The fourth column (INR fitting and registration without fine-tuning) is discussed in Section~\ref{sec:ablation}. Across all views, the proposed method recovers the right-anterior lesion, demonstrating that fine-tuning adapts the template prior to subject-specific anatomy even under more severe through-plane information loss and outside the training distribution. Reconstruction is driven by the per-subject data-consistency loss in \eqref{eqn:LR_training_objective}, with the template acting only as initialization and regularization. Because the template is healthy and, unlike supervised multi-subject models, there is no lesion-bearing training corpus from which anatomy could leak, recovered pathology can only originate from the subject's own signal. This remains a single qualitative case, however, and quantitative and expert validation on pathological cohorts is important future work.

\subsection{Ablation Study}
\label{sec:ablation}
\begin{table*}[htbp]
\caption[Ablation study of downsampling operator]{Quantitative comparison of the proposed downsampling operator versus single-point evaluation at transformed voxel centers across different fine-tuning checkpoint durations. Percentages indicate relative change from the proposed method.}
\renewcommand{\arraystretch}{1.2}
\centering
\footnotesize
\begin{tabular}{c c c l l l}
\toprule
\textbf{Metric} & \textbf{Map} & \textbf{Method} & \textbf{5 min} & \textbf{10 min} & \textbf{30 min} \\
\midrule
\multirow{4}{*}{\makecell{\textbf{NRMSE} $\downarrow$}} 
& \multirow{2}{*}{FA} 
& Proposed & \textbf{0.146} & \textbf{0.139} & \textbf{0.135} \\
& & Ablation & 0.151 {\textcolor{red}{(+3.39\%)}} & 0.145 {\textcolor{red}{(+4.68\%)}} & 0.145 {\textcolor{red}{(+8.05\%)}} \\
\cmidrule(lr){2-6}
& \multirow{2}{*}{GFA} 
& Proposed & \textbf{0.105} & \textbf{0.100} & \textbf{0.098} \\
& & Ablation & 0.106 {\textcolor{red}{(+0.70\%)}} & 0.106 {\textcolor{red}{(+5.59\%)}} & 0.106 {\textcolor{red}{(+8.61\%)}} \\
\midrule
\multirow{4}{*}{\makecell{\textbf{FSIM} $\uparrow$}} 
& \multirow{2}{*}{FA} 
& Proposed & \textbf{0.602} & \textbf{0.613} & \textbf{0.625} \\
& & Ablation & 0.596 {\textcolor{red}{(-1.00\%)}} & 0.607 {\textcolor{red}{(-0.96\%)}} & 0.616 {\textcolor{red}{(-1.41\%)}} \\
\cmidrule(lr){2-6}
& \multirow{2}{*}{GFA} 
& Proposed & \textbf{0.574} & \textbf{0.583} & \textbf{0.591} \\
& & Ablation & 0.559 {\textcolor{red}{(-2.63\%)}} & 0.561 {\textcolor{red}{(-3.82\%)}} & 0.564 {\textcolor{red}{(-4.48\%)}} \\

\midrule
\multirow{4}{*}{\makecell{\textbf{PSNR (dB)} $\uparrow$}} 
& \multirow{2}{*}{FA} 
& Proposed & \textbf{22.767} & \textbf{23.256} & \textbf{23.604} \\
& & Ablation & 22.535 {\textcolor{red}{(-1.02\%)}} & 22.900 {\textcolor{red}{(-1.53\%)}} & 22.863 {\textcolor{red}{(-3.14\%)}} \\
\cmidrule(lr){2-6}
& \multirow{2}{*}{GFA} 
& Proposed & \textbf{22.265} & \textbf{22.341} & \textbf{22.409} \\
& & Ablation & 21.960 {\textcolor{red}{(-1.37\%)}} & 22.043 {\textcolor{red}{(-1.34\%)}} & 22.015 {\textcolor{red}{(-1.75\%)}} \\
\midrule
\multirow{4}{*}{\makecell{\textbf{SSIM} $\uparrow$}} 
& \multirow{2}{*}{FA} 
& Proposed & \textbf{0.803} & \textbf{0.814} & \textbf{0.826} \\
& & Ablation & 0.796 {\textcolor{red}{(-0.83\%)}} & 0.805 {\textcolor{red}{(-1.12\%)}} & 0.805 {\textcolor{red}{(-2.50\%)}} \\
\cmidrule(lr){2-6}
& \multirow{2}{*}{GFA} 
& Proposed & \textbf{0.737} & \textbf{0.739} & \textbf{0.743} \\
& & Ablation & 0.729 {\textcolor{red}{(-1.13\%)}} & 0.730 {\textcolor{red}{(-1.13\%)}} & 0.727 {\textcolor{red}{(-2.04\%)}} \\
\bottomrule
\end{tabular}%
\label{tab:ds_ablation}
\end{table*}

We validate the three key components of the proposed framework.
\par
\noindent{\textbf{Registration and template prior}}: Removing the template initialization is equivalent to training the INR from scratch on the subject's low-resolution data alone. Table~\ref{tbl:metrics} provides this comparison, where NODF and NODF-HashEnc are both optimized from scratch.
\par
\noindent{\textbf{Fine-tuning}}: The fourth column in Figure~\ref{fig:lesion} shows the reconstruction obtained by querying the registered template INR in the subject coordinates without performing Stage 3 (Section~\ref{ssec:stage_3}). The absence of the subject-specific lesion in that column and its recovery in the fifth column of the same figure after fine-tuning demonstrate that Stage 3 is essential for adapting the prior to individual anatomy.
\par
\noindent{\textbf{Downsampling operator}}: To assess the Monte Carlo integration over the deformed voxel footprint $\tilde{\Omega}_i$ in Equation~\eqref{eqn:ds_operator}, we compare it against a variant that replaces the integration with single-point evaluation at each transformed voxel center during fine-tuning. Table~\ref{tab:ds_ablation} reports results on a randomly selected HCP scan using the same template as in Section~\ref{ssec:quantitative_and_qualitative_evaluation}. Removing the downsampling operator consistently degrades all metrics across checkpoints, with NRMSE increasing by up to 8.6\% and FSIM decreasing by up to 4.5\% on GFA at 30 minutes, confirming that explicitly modeling the spatial extent of each thick slice is important for accurate reconstruction.

\section{Discussion}
\label{sec:discussion}

The experiments indicate that the hash-grid encodings~\cite{muller2022instant} adopted in NODF-HashEnc~\cite{dwedari2024estimating} limit the retrieval of missing through-plane details in thick slices. Gradient updates primarily affect grid entries at supervised positions. However, at inference time, off-grid positions rely on trilinear interpolation of local embeddings that, lacking encoding of global spatial relationships, fail to maintain coherent microstructural features across wide interslice gaps, yielding diminished anisotropy and directional information in the reconstructed slices and degraded tractography with reduced streamline density and early tract termination. In contrast, NODF~\cite{consagra2024neural} with SIREN~\cite{sitzmann2020implicit} activations, although slow to train, learns a globally coherent, continuous spatial basis that captures spatial correlations across the full volume. This global coherence, combined with the proposed prior, enables effective super-resolution under extreme anisotropy.

The quantitative results further expose a discrepancy between some image quality metrics 
and perceptual quality: NODF initially achieves nominally better PSNR and SSIM 
scores despite producing oversmoothed images, and the metrics counterintuitively 
decrease or fluctuate as training progresses, a behavior explained by the perception-distortion tradeoff~\cite{blau2018perception}. Qualitative inspection confirms NODF's reconstruction remains blurry even after 1 hour, whereas our method produces sharper outputs with more anatomical detail. PSNR and SSIM, developed for natural images, have well-documented limitations in medical imaging: they systematically favor blurred reconstructions over those that preserve fine anatomical detail~\cite{breger2025study} and correlate poorly with the assessments of expert radiologists~\cite{mason2019comparison}. On domain-specific metrics directly relevant to diffusion MRI analysis, including ODF reconstruction error, eigenvector angular error, and tractography fidelity, our method outperforms both baselines across all training times considered.

\section{Conclusion}
\label{conclusion}

We propose a template-guided implicit neural representation (INR) framework for diffusion MRI super-resolution. Using transfer learning, we pre-train an INR on a high-resolution template and map it to new subjects via deformable registration, serving as a prior for subsequent subject-specific fine-tuning. Evaluated against two recent INR-based methods, our framework achieves significantly faster training and higher-quality reconstructions on Human Connectome Project diffusion data, recovering fine anatomical structures that neither baseline could resolve at \(4\times\) through-plane super-resolution. Future work could adopt registration models trained specifically for dMRI-derived scalar images, where improved spatial mapping fidelity is expected to benefit reconstruction quality; extend to multiple b-values and angular super-resolution; incorporate measured slice-excitation profiles into the downsampling operator; and pursue quantitative and expert validation on pathological cohorts.

\section*{Acknowledgements}
Research reported in this manuscript was supported by the National Institute of Neurological Disorders and Stroke, the National Institute of Mental Health, and the National Institute of Biomedical Imaging and Bioengineering of the National Institutes of Health under Award Numbers R01NS125307, R01MH132610, R01MH125860, R01EB032378, and R01MH119222. The content is solely the responsibility of the authors and does not necessarily represent the official views of the National Institutes of Health. Data were provided by the Human Connectome Project, WU-Minn Consortium (Principal Investigators: David Van Essen and Kamil Ugurbil; 1U54MH091657), funded by the 16 NIH Institutes and Centers that support the NIH Blueprint for Neuroscience Research; and by the McDonnell Center for Systems Neuroscience at Washington University.



\bibliography{bibliography}

\begin{thebibliography}{29}
\providecommand{\natexlab}[1]{#1}
\providecommand{\url}[1]{\texttt{#1}}
\expandafter\ifx\csname urlstyle\endcsname\relax
  \providecommand{\doi}[1]{doi: #1}\else
  \providecommand{\doi}{doi: \begingroup \urlstyle{rm}\Url}\fi

\bibitem[Alexander et~al.(2017)Alexander, Zikic, Ghosh, Tanno, Wottschel, Zhang, Kaden, Dyrby, Sotiropoulos, Zhang, et~al.]{alexander2017image}
Daniel~C Alexander, Darko Zikic, Aurobrata Ghosh, Ryutaro Tanno, Viktor Wottschel, Jiaying Zhang, Enrico Kaden, Tim~B Dyrby, Stamatios~N Sotiropoulos, Hui Zhang, et~al.
\newblock Image quality transfer and applications in diffusion mri.
\newblock \emph{NeuroImage}, 152:\penalty0 283--298, 2017.

\bibitem[Avants et~al.(2014)Avants, Tustison, Stauffer, Song, Wu, and Gee]{avants2014insight}
Brian~B Avants, Nicholas~J Tustison, Michael Stauffer, Gang Song, Baohua Wu, and James~C Gee.
\newblock The insight toolkit image registration framework.
\newblock \emph{Frontiers in neuroinformatics}, 8:\penalty0 44, 2014.

\bibitem[Billah and Bouix(2020)]{luigipnlpipe}
Tashrif Billah and Sylvain Bouix.
\newblock A {Luigi} workflow joining individual modules of an {MRI} processing pipeline.
\newblock \url{https://github.com/pnlbwh/luigi-pnlpipe}, 2020.
\newblock DOI: 10.5281/zenodo.3666802.

\bibitem[Blau and Michaeli(2018)]{blau2018perception}
Yochai Blau and Tomer Michaeli.
\newblock The perception-distortion tradeoff.
\newblock In \emph{Proceedings of the IEEE conference on computer vision and pattern recognition}, pages 6228--6237, 2018.

\bibitem[Breger et~al.(2025)Breger, Biguri, Landman, Selby, Amberg, Brunner, Gr{\"o}hl, Hatamikia, Karner, Ning, et~al.]{breger2025study}
Anna Breger, Ander Biguri, Malena~Sabat{\'e} Landman, Ian Selby, Nicole Amberg, Elisabeth Brunner, Janek Gr{\"o}hl, Sepideh Hatamikia, Clemens Karner, Lipeng Ning, et~al.
\newblock A study of why we need to reassess full reference image quality assessment with medical images.
\newblock \emph{Journal of Imaging Informatics in Medicine}, pages 1--26, 2025.

\bibitem[Chen et~al.(2021)Chen, Liu, and Wang]{chen2021learning}
Yinbo Chen, Sifei Liu, and Xiaolong Wang.
\newblock Learning continuous image representation with local implicit image function.
\newblock In \emph{Proceedings of the IEEE/CVF conference on computer vision and pattern recognition}, pages 8628--8638, 2021.

\bibitem[Consagra et~al.(2024)Consagra, Ning, and Rathi]{consagra2024neural}
William Consagra, Lipeng Ning, and Yogesh Rathi.
\newblock Neural orientation distribution fields for estimation and uncertainty quantification in diffusion mri.
\newblock \emph{Medical Image Analysis}, 93:\penalty0 103105, 2024.

\bibitem[Descoteaux et~al.(2007)Descoteaux, Angelino, Fitzgibbons, and Deriche]{descoteaux2007regularized}
Maxime Descoteaux, Elaine Angelino, Shaun Fitzgibbons, and Rachid Deriche.
\newblock Regularized, fast, and robust analytical q-ball imaging.
\newblock \emph{Magnetic Resonance in Medicine: An Official Journal of the International Society for Magnetic Resonance in Medicine}, 58\penalty0 (3):\penalty0 497--510, 2007.

\bibitem[Dwedari et~al.(2024)Dwedari, Consagra, M{\"u}ller, Turgut, Rueckert, and Rathi]{dwedari2024estimating}
Mohammed~Munzer Dwedari, William Consagra, Philip M{\"u}ller, {\"O}zg{\"u}n Turgut, Daniel Rueckert, and Yogesh Rathi.
\newblock Estimating neural orientation distribution fields on high resolution diffusion mri scans.
\newblock In \emph{International Conference on Medical Image Computing and Computer-Assisted Intervention}, pages 307--317. Springer, 2024.

\bibitem[Elaldi et~al.(2021)Elaldi, Dey, Kim, and Gerig]{elaldi2021equivariant}
Axel Elaldi, Neel Dey, Heejong Kim, and Guido Gerig.
\newblock Equivariant spherical deconvolution: Learning sparse orientation distribution functions from spherical data.
\newblock In \emph{International Conference on Information Processing in Medical Imaging}, pages 267--278. Springer, 2021.

\bibitem[Garyfallidis et~al.(2014)Garyfallidis, Brett, Amirbekian, Rokem, Van Der~Walt, Descoteaux, Nimmo-Smith, and Contributors]{garyfallidis2014dipy}
Eleftherios Garyfallidis, Matthew Brett, Bagrat Amirbekian, Ariel Rokem, Stefan Van Der~Walt, Maxime Descoteaux, Ian Nimmo-Smith, and Dipy Contributors.
\newblock Dipy, a library for the analysis of diffusion mri data.
\newblock \emph{Frontiers in neuroinformatics}, 8:\penalty0 8, 2014.

\bibitem[Glasser et~al.(2013)Glasser, Sotiropoulos, Wilson, Coalson, Fischl, Andersson, Xu, Jbabdi, Webster, Polimeni, et~al.]{glasser2013minimal}
Matthew~F Glasser, Stamatios~N Sotiropoulos, J~Anthony Wilson, Timothy~S Coalson, Bruce Fischl, Jesper~L Andersson, Junqian Xu, Saad Jbabdi, Matthew Webster, Jonathan~R Polimeni, et~al.
\newblock The minimal preprocessing pipelines for the human connectome project.
\newblock \emph{Neuroimage}, 80:\penalty0 105--124, 2013.

\bibitem[Hendriks et~al.(2025)Hendriks, Vilanova, and Chamberland]{hendriks2025implicit}
Tom Hendriks, Anna Vilanova, and Maxime Chamberland.
\newblock Implicit neural representation of multi-shell constrained spherical deconvolution for continuous modeling of diffusion mri.
\newblock \emph{Imaging Neuroscience}, 3:\penalty0 imag\_a\_00501, 2025.

\bibitem[Jo et~al.(2025)Jo, Joo, Choi, Joe, and Lee]{jo2025white}
Young~Tak Jo, Sung~Woo Joo, Woohyeok Choi, Soohyun Joe, and Jungsun Lee.
\newblock White matter tract alterations in schizophrenia identified by dti-based probabilistic tractography: a multisite harmonisation study.
\newblock \emph{Acta Neuropsychiatrica}, 37:\penalty0 e47, 2025.

\bibitem[Lyon et~al.(2023)Lyon, Armitage, and {\'A}lvarez]{lyon2023spatio}
Matthew Lyon, Paul Armitage, and Mauricio~A {\'A}lvarez.
\newblock Spatio-angular convolutions for super-resolution in diffusion mri.
\newblock \emph{Advances in Neural Information Processing Systems}, 36:\penalty0 12457--12475, 2023.

\bibitem[Mason et~al.(2019)Mason, Rioux, Clarke, Costa, Schmidt, Keough, Huynh, and Beyea]{mason2019comparison}
Allister Mason, James Rioux, Sharon~E Clarke, Andreu Costa, Matthias Schmidt, Valerie Keough, Thien Huynh, and Steven Beyea.
\newblock Comparison of objective image quality metrics to expert radiologists’ scoring of diagnostic quality of mr images.
\newblock \emph{IEEE transactions on medical imaging}, 39\penalty0 (4):\penalty0 1064--1072, 2019.

\bibitem[Mildenhall et~al.(2021)Mildenhall, Srinivasan, Tancik, Barron, Ramamoorthi, and Ng]{mildenhall2021nerf}
Ben Mildenhall, Pratul~P Srinivasan, Matthew Tancik, Jonathan~T Barron, Ravi Ramamoorthi, and Ren Ng.
\newblock Nerf: Representing scenes as neural radiance fields for view synthesis.
\newblock \emph{Communications of the ACM}, 65\penalty0 (1):\penalty0 99--106, 2021.

\bibitem[M{\"u}ller and Kassubek(2024)]{muller2024toward}
Hans-Peter M{\"u}ller and Jan Kassubek.
\newblock Toward diffusion tensor imaging as a biomarker in neurodegenerative diseases: technical considerations to optimize recordings and data processing.
\newblock \emph{Frontiers in Human Neuroscience}, 18:\penalty0 1378896, 2024.

\bibitem[M{\"u}ller et~al.(2022)M{\"u}ller, Evans, Schied, and Keller]{muller2022instant}
Thomas M{\"u}ller, Alex Evans, Christoph Schied, and Alexander Keller.
\newblock Instant neural graphics primitives with a multiresolution hash encoding.
\newblock \emph{ACM transactions on graphics (TOG)}, 41\penalty0 (4):\penalty0 1--15, 2022.

\bibitem[Park et~al.(2019)Park, Florence, Straub, Newcombe, and Lovegrove]{park2019deepsdf}
Jeong~Joon Park, Peter Florence, Julian Straub, Richard Newcombe, and Steven Lovegrove.
\newblock Deepsdf: Learning continuous signed distance functions for shape representation.
\newblock In \emph{Proceedings of the IEEE/CVF conference on computer vision and pattern recognition}, pages 165--174, 2019.

\bibitem[Sitzmann et~al.(2020)Sitzmann, Martel, Bergman, Lindell, and Wetzstein]{sitzmann2020implicit}
Vincent Sitzmann, Julien Martel, Alexander Bergman, David Lindell, and Gordon Wetzstein.
\newblock Implicit neural representations with periodic activation functions.
\newblock \emph{Advances in neural information processing systems}, 33:\penalty0 7462--7473, 2020.

\bibitem[Sotiropoulos et~al.(2013)Sotiropoulos, Jbabdi, Xu, Andersson, Moeller, Auerbach, Glasser, Hernandez, Sapiro, Jenkinson, et~al.]{sotiropoulos2013advances}
Stamatios~N Sotiropoulos, Saad Jbabdi, Junqian Xu, Jesper~L Andersson, Steen Moeller, Edward~J Auerbach, Matthew~F Glasser, Moises Hernandez, Guillermo Sapiro, Mark Jenkinson, et~al.
\newblock Advances in diffusion mri acquisition and processing in the human connectome project.
\newblock \emph{Neuroimage}, 80:\penalty0 125--143, 2013.

\bibitem[Spears and Fletcher(2024)]{spears2024learning}
Tyler Spears and P~Thomas Fletcher.
\newblock Learning spatially-continuous fiber orientation functions.
\newblock In \emph{2024 IEEE International Symposium on Biomedical Imaging (ISBI)}, pages 1--5. IEEE, 2024.

\bibitem[Tanno et~al.(2017)Tanno, Worrall, Ghosh, Kaden, Sotiropoulos, Criminisi, and Alexander]{tanno2017bayesian}
Ryutaro Tanno, Daniel~E Worrall, Aurobrata Ghosh, Enrico Kaden, Stamatios~N Sotiropoulos, Antonio Criminisi, and Daniel~C Alexander.
\newblock Bayesian image quality transfer with cnns: exploring uncertainty in dmri super-resolution.
\newblock In \emph{International Conference on Medical Image Computing and Computer-Assisted Intervention}, pages 611--619. Springer, 2017.

\bibitem[Tian et~al.(2024)Tian, Greer, Kwitt, Vialard, San Jos{\'e}~Est{\'e}par, Bouix, Rushmore, and Niethammer]{tian2024unigradicon}
Lin Tian, Hastings Greer, Roland Kwitt, Francois-Xavier Vialard, Ra{\'u}l San Jos{\'e}~Est{\'e}par, Sylvain Bouix, Richard Rushmore, and Marc Niethammer.
\newblock unigradicon: A foundation model for medical image registration.
\newblock In \emph{International Conference on Medical Image Computing and Computer-Assisted Intervention}, pages 749--760. Springer, 2024.

\bibitem[Tournier et~al.(2008)Tournier, Yeh, Calamante, Cho, Connelly, and Lin]{tournier2008resolving}
J-Donald Tournier, Chun-Hung Yeh, Fernando Calamante, Kuan-Hung Cho, Alan Connelly, and Ching-Po Lin.
\newblock Resolving crossing fibres using constrained spherical deconvolution: validation using diffusion-weighted imaging phantom data.
\newblock \emph{Neuroimage}, 42\penalty0 (2):\penalty0 617--625, 2008.

\bibitem[Van~Essen et~al.(2013)Van~Essen, Smith, Barch, Behrens, Yacoub, Ugurbil, Consortium, et~al.]{van2013wu}
David~C Van~Essen, Stephen~M Smith, Deanna~M Barch, Timothy~EJ Behrens, Essa Yacoub, Kamil Ugurbil, Wu-Minn~HCP Consortium, et~al.
\newblock The wu-minn human connectome project: an overview.
\newblock \emph{Neuroimage}, 80:\penalty0 62--79, 2013.

\bibitem[Wu et~al.(2024)Wu, Cheng, Li, Zou, Yang, Fan, Liang, and Wang]{wu2024csr}
Ruoyou Wu, Jian Cheng, Cheng Li, Juan Zou, Jing Yang, Wenxin Fan, Yong Liang, and Shanshan Wang.
\newblock Csr-dmri: Continuous super-resolution of diffusion mri with anatomical structure-assisted implicit neural representation learning.
\newblock In \emph{International Workshop on Machine Learning in Medical Imaging}, pages 114--123. Springer, 2024.

\bibitem[Zhao et~al.(2025)Zhao, You, Ren, Ji, Liu, and Lu]{zhao2025application}
Xinle Zhao, Mengyue You, Wenyu Ren, Lixin Ji, Yongbo Liu, and Meng Lu.
\newblock The application of diffusion tensor imaging in patients with mild cognitive impairment: a systematic review and meta-analysis.
\newblock \emph{Frontiers in Neurology}, 16:\penalty0 1467578, 2025.

\end{thebibliography}
\end{document}